\documentclass{article}
\usepackage{natbib}
\usepackage{graphicx}
\usepackage{tabularx}
\usepackage{amsmath}
\usepackage{float}
\usepackage{booktabs}
\usepackage[left]{lineno} % line numbers
\usepackage{authblk} % for proper author affiliations display
\usepackage{threeparttable} % for notes under tables. No idea why that's not part of the default functionality.
\usepackage{multicol}
\usepackage{tabularray}

\usepackage{lipsum}
\usepackage{background}
\backgroundsetup{position=current page.east, angle=-90, nodeanchor=east, vshift=-5mm, opacity=1, scale=2.5,
    contents=Preprint -- under review}

\usepackage[font=bf]{caption} % Makes figure caption bold
\usepackage[top=2.5cm,right=2.5cm, left=2.5cm, bottom=2.5cm]{geometry} % page margins
\usepackage{titlesec}
\titleformat*{\subsection}{\normalsize\bfseries}

\usepackage{geometry}
\usepackage{hyperref}
\hypersetup{
    colorlinks=true,
    citecolor=black, 
    linkcolor=black,
    filecolor=black,      
    urlcolor=blue,
    }
\usepackage{cleveref}
\usepackage[superscript,biblabel]{cite}
\begin{document}
%\linenumbers 
\widowpenalty10000 % to prevent single/ half lines on new pages
% textwidth in cm: \printinunitsof{cm}\prntlen{\textwidth} 16.04 cm

% Title and author details. 
\title{Global coal trade is resilient to maritime chokepoints}
\author[1,*]{Jorrit Gosens}
\author[1,2]{Alex B.H. Turnbull}
\author[1]{Frank Jotzo}
\affil[1]{\footnotesize Centre for Climate and Energy Policy, Crawford School of Public Policy, Australian National University, Acton, 2600 Australian Capital Territory, Australia}
\affil[2]{\footnotesize Sagax Capital, 048583 Singapore}
\affil[*]{\footnotesize Corresponding author; jorrit.gosens@anu.edu.au}
\date{} % exclude date below title

\maketitle
\renewcommand{\abstractname}{Abstract}
\begin{abstract}
Maritime chokepoints and their potential disruption of global trade in energy find renewed attention. We analyse global trade in coal, and find that trade volumes and prices are highly resilient to maritime chokepoints. Feasible chokepoints do not truly sever supply from the seaborne market. Effects on costs and revenues are further moderated by relatively large potential for re-routing of bilateral trade flows, with countries switching to alternative suppliers or consumers. We assess costs to importers would rise by as little as 0.5 \$/t or less in case of closures of most feasible chokepoints. The exception is a restriction to maritime traffic in the South and East China Sea, which could raise costs by 10 \$/t for China, whilst reducing costs for other importers in the region by similar levels. Maritime chokepoints do create geographical separation of regional markets, with differentiated effects on costs to importers and revenues to exporters in different regions.
\end{abstract}

%\subsection*{Keywords}
%Maritime chokepoints, coal trade, energy security, linear optimization

\section{Introduction}\label{introduction}
The Hormuz Strait crisis has put renewed attention on maritime chokepoints and their potential effects on global energy markets, with much concern about effects on energy prices\cite{IEA2026Strait}, and resulting effects on inflation and exacerbation of the cost of living crisis\cite{Mishra2026How,Semieniuk2025Best,Neri2023Energy}. The blockage of the Suez canal, piracy and militant activity in the Bab el-Mandeb Strait, and drought-induced restrictions to traffic through the Panama canal in recent years have previously highlighted the potential disruption to global trade of maritime chokepoints\cite{UNCTAD2024Navigating,Pratson2023Assessing,S2023Panama,Ulrichsen2026Maritime,Verschuur2025Systemic}.

Analysis of how energy costs might respond to changes in energy infrastructure and production has seen contributions from the field of operations research. A focal topic in this literature has been cost-optimal investment in transmission (and generation) infrastructure required to enable decarbonisation of power systems\cite{Brown2018Synergies,Parzen2023PyPSA}. Examples focused on coal markets include analyses of Chinese investment in railway and transmission networks and resulting changes to global coal trade patterns\cite{Gosens2022Chinas,Paulus2011Coal}. Effects of disruptions to existing energy infrastructure or transport routes, however, have not been a focus of this literature to date.

Specific analysis of effects of maritime chokepoints on global trade include an 2011 analysis that described effects of disruptions to transport routes for oil on e.g., GDP, spending, and inflation, using input-output analysis\cite{Komiss2011Economic} and the simplifying assumption that increases in oil prices in importing countries are directly related to their share of imports normally routed via particular maritime chokepoints. More recent studies have analysed how closures of different maritime chokepoints might re-route trade flows, and subsequently lead to delays in shipping, and reductions in seaport activity\cite{UNCTAD2024Navigating,Pratson2023Assessing}, or how the costs of imports might rise as transport costs for goods normally arriving via different maritime chokepoints are re-routed\cite{Verschuur2025Systemic}. Each of these analyses assumes that countries would continue to source their imports from the same exporter; an assumption that may hold to some or large extent for manufactured goods, but not likely for fungible commodities like fossil fuels. A recent analysis specifically on the effects of the Hormuz strait closure on global LNG trade does consider and report on how trade flows may be re-routed but does not report effects on costs to importers or revenues for exporters\cite{Meza2026Implications}.

The analysis reported here is the first to use a source-and-sink cost-optimisation model to analyse effects of disruption or closure of maritime chokepoints on global markets for coal. Our model contains representations of global maritime trade networks, coal mines, coal demand centres, export and import terminals and land-based transport with unparalleled detail. We report on resulting re-routing of trade flows, and effects on costs to importers and revenues for exporters. The results suggest that the global coal market is surprisingly resilient to potential disruptions from maritime chokepoints. Whilst bilateral trade patterns may change significantly, disruptions at the world’s major maritime chokepoints would increase import prices only to a very small extent. The exception is the Taiwan Strait and South and East China seas, where partial closure which would cause a significant substitution from seaborne imports to domestically mined coal in China. The effects on cost to importers and revenues for exporters are not uniform across the globe, because of geographical separation of regional markets that results from chokepoints.

\section{Results}\label{results}
We analyse effects on coal trade of disruptions at the following maritime chokepoints: 1) the Panama Canal, 2) the Suez Canal, 3) the Bosporus or Danish Strait, 4) the Straits of Malacca and Indonesia, and 5) the Taiwan Strait, and the South and East China Sea. The explanation for this selection a plot locating these chokepoints on a world map (Fig. \ref{fig:map_chokepoints}) are provided in the Methods section.

\subsection{Panama Canal restrictions}\label{panama-canal-restrictions}
A closure of the Panama Canal would primarily restrict existing trade flows of coal from Colombia and the US to Asia. Colombian mines supplying thermal coal are predominantly located in Colombia's Northern provinces and export via ports on the Caribbean coast, i.e., East of the Panama Canal. US coal mines, which supply both thermal coal and high-grade coking coal into the seaborne market, are highly concentrated in the Appalachian basin, with coal exported via ports in Virginia, on the US East coast. Canada also exports high-grade coking coal to Asia, but this is mined mostly in Alberta and British Colombia, and exported via ports in Vancouver on Canada's west coast, meaning they would be unaffected by a closure of the Panama Canal.

\begin{figure}[H] % hbtp for here, top, bottom, next available page. H for really just here
    \centering
    \includegraphics[width=1.0\textwidth]{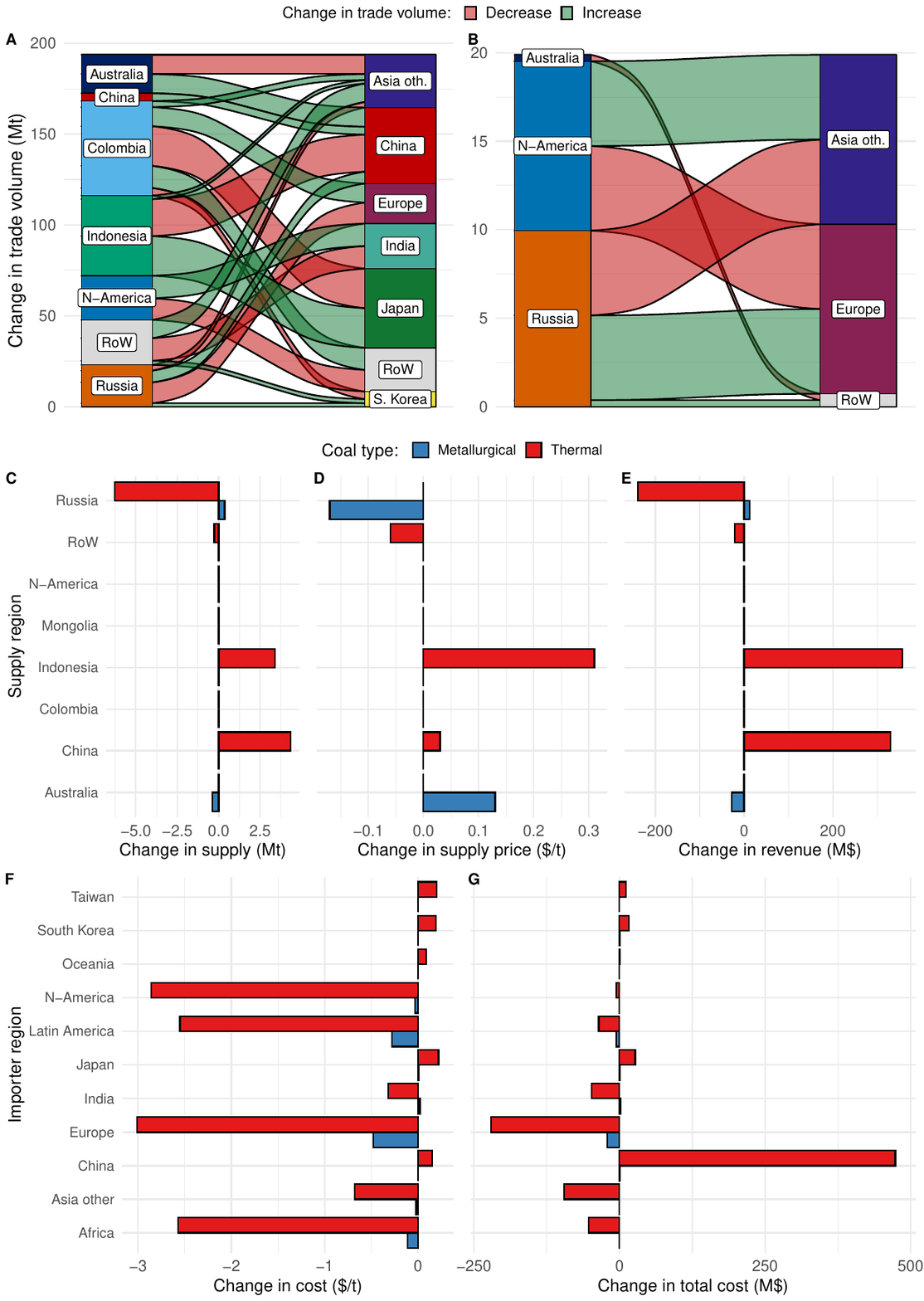}
        \caption{Changes to global coal trade in case of a restriction to maritime traffic through the Panama Canal. 
        \textmd{Changes in volumes of for thermal coal (A) and metallurgical coal (B) traded between key exporter and importer country (groups), supply volume (C), price (D) and revenue (E) for exporters, and cost per ton (F) and total costs (G) for importers. Note: trade flow changes $<$1 Mt in panel A censored for legibility.}
        }
    \label{fig:panama}
\end{figure}

In our baseline projection, our model suggests 26.4 Mt of Colombian coal and 9.9 Mt of US coal is transported to markets in Asia via the Panama Canal. In a scenario where transport via the Panama Canal is restricted, Colombian exports of thermal coal originally bound for Japan and South Korea, are instead redirected to Europe, Northern Africa, and Turkey (Fig.~\ref{fig:panama}).

This precipitates a further reshuffling of suppliers, as US thermal coal supplies to Northern Africa and Turkey are now redirected to India. This displaces Indian imports from South Africa, which are now redirected to other Asian markets. The supply of thermal coal from Colombia originally bound for Japan is filled by Indonesian suppliers instead, who reduce supplies to China. That gap is in turn filled by Australian and Chinese domestic suppliers. US thermal coal shipments originally bound for northern provinces of China via the Panama Canal are still exported to China, but to more southern provinces, and via the Suez Canal.

Our baseline projection further suggests a minimal 0.35 Mt of Australian metallurgical coal makes its way via the Panama Canal to Latin and North American destinations. In the case of a closure of the canal, these markets are supplied by Russia from ports on the Black Sea instead. This marginally lifts the cost of Russian supplies to Turkey, which switches out 4.8 Mt of metallurgical coal imports from Russia to suppliers from the US instead, which is offset by an exact opposite switch between these two suppliers to European markets.

We note that the resulting increase in supplies of metallurgical coal from Russia, whilst very small, has a counterintuitive effect on prices for Russian coal. This is because we summarize results for three types of metallurgical coal here, hard coking coal (HCC), semi-soft coking coal (SSCC), and pulverised coal for injection (PCI), which have distinctly different prices in international coal markets. The small increase in Russian metallurgical coal supplies is entirely SSCC, suppressing the average price per ton for the mix of metallurgical coal it exports (Suppl. Table 1). An opposite effect holds for Australian metallurgical coal exports.

Overall, the effect of a closure of the Panama Canal leads to cost reductions in markets for coal on the Atlantic Ocean, and increases in
the Asia-Pacific region, as supplies from Colombia and the US East coast are redirected Eastwards.

The more surprising result is how minimal the changes to global coal markets in the case of a closure of the Panama Canal are projected to be.

Despite about 36 Mt of coal transport through the canal in our baseline scenario, and despite the resulting rerouting of about 100 Mt of thermal coal and 10 Mt of metallurgical coal between different suppliers and consumers (Fig.~\ref{fig:panama}A\&B), total changes to supplies from different countries amount to a circa 6 Mt drop in Russian exports, and increases of 3 Mt from Indonesia and China (Fig.~\ref{fig:panama}C). The net effect on the two suppliers exporting via the Panama Canal, Colombia and the US, is effectively zero: total exported volumes, prices fetched, and total revenue, are entirely unchanged in this scenario for these countries.

Results on costs are small as well: costs for thermal coal in the Asia-Pacific region amount to as little as an 0.2\% cost increase for thermal coal; whilst costs in smaller markets on the Atlantic are reduced by about 3 to 4\% (Fig.~\ref{fig:panama}F, Suppl. Table 2). Changes in metallurgical coal costs, as well as prices fetched by suppliers for both thermal and metallurgical, are entirely negligible (Fig.~\ref{fig:panama}D\&F).

We should note that the volume of coal transported through the Panama Canal in our baseline scenario is relatively high versus reported volumes of around 15 Mt for the years 2021-2023\cite{Canal2023Principal}, and volumes of as little as 3 Mt in 2024 and 6 Mt in 2025\cite{Canal2026Estadisticas}. It has been well documented that traffic via the Panama canal has already been restricted since a number of years, due to climate-change induced drought, however\cite{S2023Panama,Council2023Whats}, suggesting that this chokepoint has already materialised to an extent. This means both that our results presented in Fig.~\ref{fig:panama}, whilst relatively minor, may still be an exaggeration of real-world effects in the case of a Panama Canal closure, and effects may be smaller still when considering the current relatively restricted status of the Panama Canal as the baseline.

\subsection{Suez Canal restrictions}\label{suez-canal-restrictions}
In our baseline scenario, 1.4 Mt of thermal coal is transported via the Suez Canal from Russian ports on the Black Sea to China, and 15.0 Mt of metallurgical coal from the same Russian ports, the bulk of it bound for India, with the remainder bound for other Asian markets.

\begin{figure}[H] % hbtp for here, top, bottom, next available page. H for really just here
    \centering
    \includegraphics[width=1.0\textwidth]{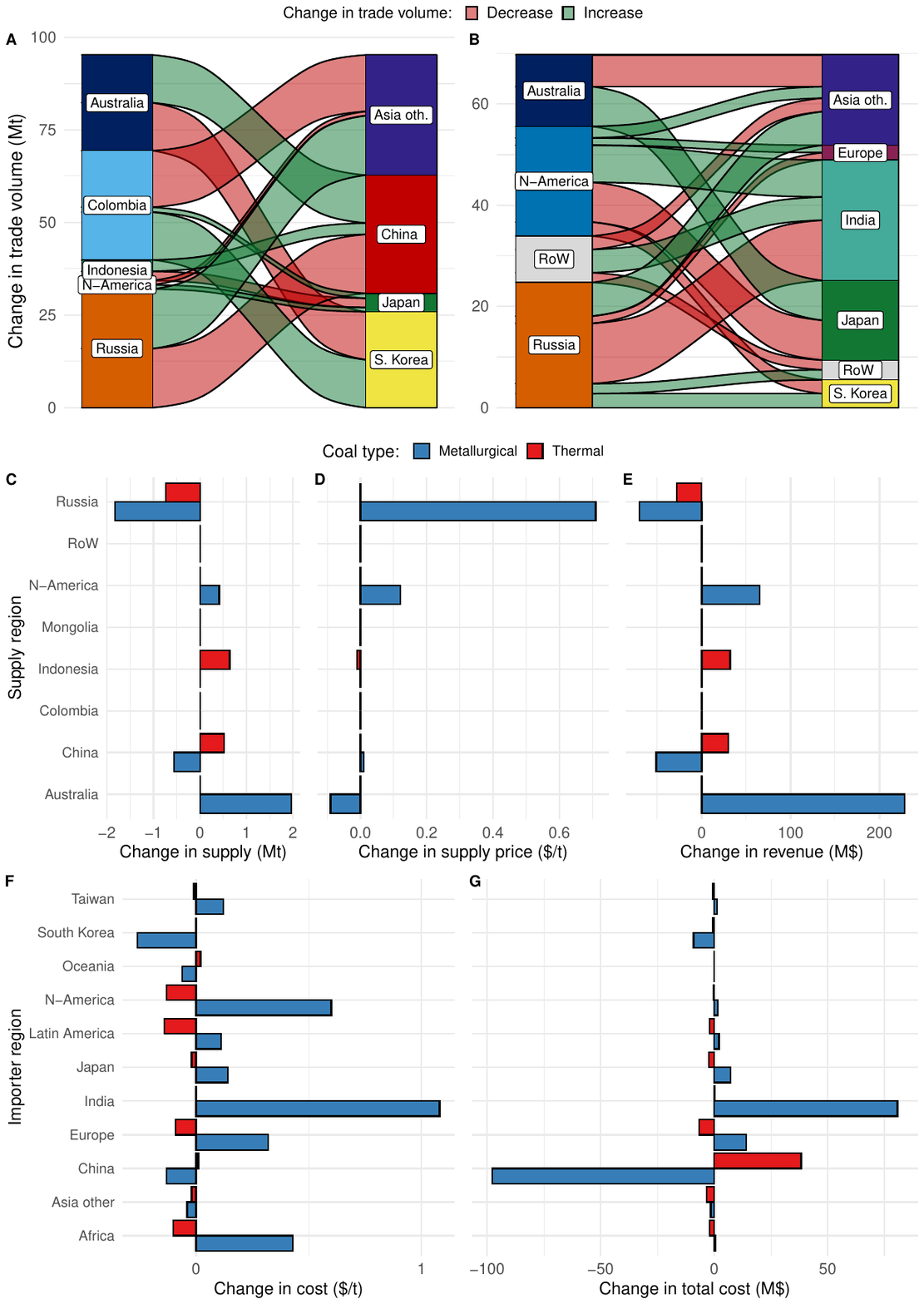}
        \caption{Changes to global coal trade in case of a restriction to maritime traffic through the Suez Canal. 
        \textmd{Changes in volumes of for thermal coal (A) and metallurgical coal (B) traded between key exporter and importer country (groups), supply volume (C), price (D) and revenue (E) for exporters, and cost per ton (F) and total costs (G) for importers. Note: trade flow changes $<$1 Mt in panel A and changes $<$0.5 Mt in panel B censored for legibility.}
        }
    \label{fig:suez}
\end{figure}

In a scenario where transport via the Suez Canal is restricted, the decrease of exports of metallurgical coal from ports in Western Russia towards Asian destinations is nearly entirely offset by an increase from ports in Russia's far East (Fig.~\ref{fig:suez}B\&C).

This has an effect of suppressing costs to consumers in East Asia, whilst raising costs to consumers in markets on the Atlantic Ocean, and India. The differentiation between upward and downward effects in markets on either side of the Suez Canal are not as clearcut as they were in the scenario for the Panama Canal closure. This is likely because the Russian rail network provides an alternative link between markets on the Atlantic Ocean and in Asia.

Effects of a closure of the Suez Canal are moderated by the substantial resulting rerouting of trade flows between different exporters and importers. Total changes in supplies from Russia are much smaller than the volume it exported via the Suez Canal. Resulting effects on costs are limited: less than 1\% of an increase in costs of metallurgical coal in India; whilst effects on other markets and for thermal coal are even smaller (Fig. ~\ref{fig:suez}F).

Our baseline projection of coal transported via the Suesz canal is very close to the 13 Mt of Russian metallurgical reported by Kpler (a maritime shipping tracking service) reported for 2024\cite{Kpler2024Update}. The same source notes that in earlier years, a further 30 Mt of metallurgical coal from the US and Australia would have passed through the Suez Canal. We provide results for a year with trade flows that closely match such a situation in Supplementary Fig. ~\ref{fig:suez2020}; though cost increases similarly remain below 1 \$/t. Kpler notes but that these flows of US and Australian coal have since been re-routed around the Cape of Good Hope, providing further support for the idea that these trade flows are relatively easily rerouted\cite{Kpler2024Update}.

\subsection{Bosporus or Danish Strait
restrictions}\label{bosporus-or-danish-strait-restrictions}

There are Russian coal export ports with about 70 Mt of handling capacity located on the Black Sea. Alternative export routes from Western Russia are from ports in the Baltics, via the Danish Strait (ca. 45 Mt), from ports on the Barents Sea (ca. 20 Mt), and via rail lines into the EU (ca. 25 Mt), or from ports in Russia's far East (ca. 120 Mt)\cite{Wood2024Russia}. The re-routing of coal flows to such alternative export ports is limited by the capacity of rail lines within Russia.

A closure of the Bosporus Strait or limitation to maritime traffic within the Black Sea would cut off ports on the Black Sea from the global seaborne market. In our baseline scenario, a total of 23 Mt of coal is exported from Black sea ports; in a scenario where this route is restricted, total Russian exports fall by about 19.5 Mt, suggesting that this is a close equivalent to a highly critical chokepoint (see the typology of chokepoints in the Methods section) in global coal trade, as the realistic ability to redirect coal flows appears to be limited.

\begin{figure}[H] % hbtp for here, top, bottom, next available page. H for really just here
    \centering
    \includegraphics[width=1.0\textwidth]{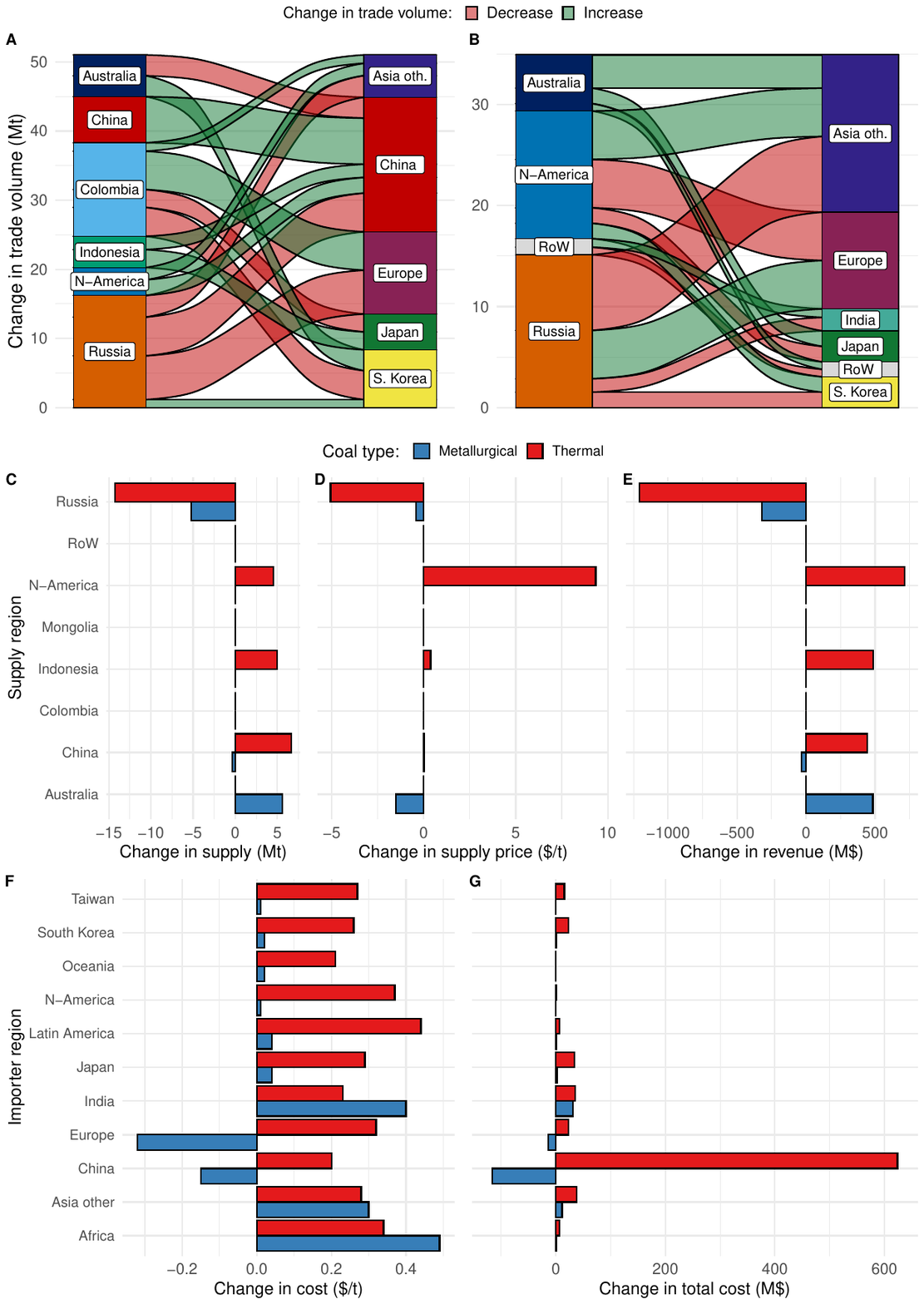}
        \caption{Changes to global coal trade in case of a restriction to maritime traffic through the Bosporus Strait. 
        \textmd{Changes in volumes of for thermal coal (A) and metallurgical coal (B) traded between key exporter and importer country (groups), supply volume (C), price (D) and revenue (E) for exporters, and cost per ton (F) and total costs (G) for importers. Note: trade flow changes $<$1 Mt in panel A and changes $<$0.5 Mt in panel B censored for legibility.}
        }
    \label{fig:blacksea}
\end{figure}

Russian supplies of thermal coal are replaced by supplies from China, Indonesia and North America in this scenario, whilst metallurgical coal supplies are replaced by Australia (Fig.~\ref{fig:blacksea}A\&B). Whilst the reduction in Russian supplies into the global seaborne market lifts costs for almost all consumers, the increases in costs, at below 0.5 \$/t (Fig.~\ref{fig:blacksea}F) are surprisingly minimal. This suggests global supply curves are still relatively flat at current levels, with limited increases in cost when supply capacity in the order of several dozen Mt is removed.

In a scenario where maritime traffic through the Danish Strait would be restricted, cutting off Russian ports in the Baltic Sea, resulting changes to global coal flows and prices show a similar pattern, though with even smaller costs increases (Supplementary Fig. \ref{fig:danish}).

\subsection{Straits of Malacca and Indonesia restrictions}\label{straits-of-malacca-and-indonesia-restrictions}
The Strait of Malacca is the primary shipping route between the Indian and Pacific Ocean, and considered the second most important chokepoint for global energy trade after the Hormuz Strait\cite{EIA2026Strait,Yin2021Energy}. There are, however, alternate though slightly longer routes between the islands of Indonesia, via the Makassar Strait, the Lombok strait, or the Sunda Strait, in case of a restriction to traffic via the Malacca Strait\cite{EIA2026Strait}. For this scenario we presume a 20\% increase in shipping costs as traffic is re-routed via one of these other straits, for all trade flows that would normally flow via these straits.

\begin{figure}[H] % hbtp for here, top, bottom, next available page. H for really just here
    \centering
    \includegraphics[width=1.0\textwidth]{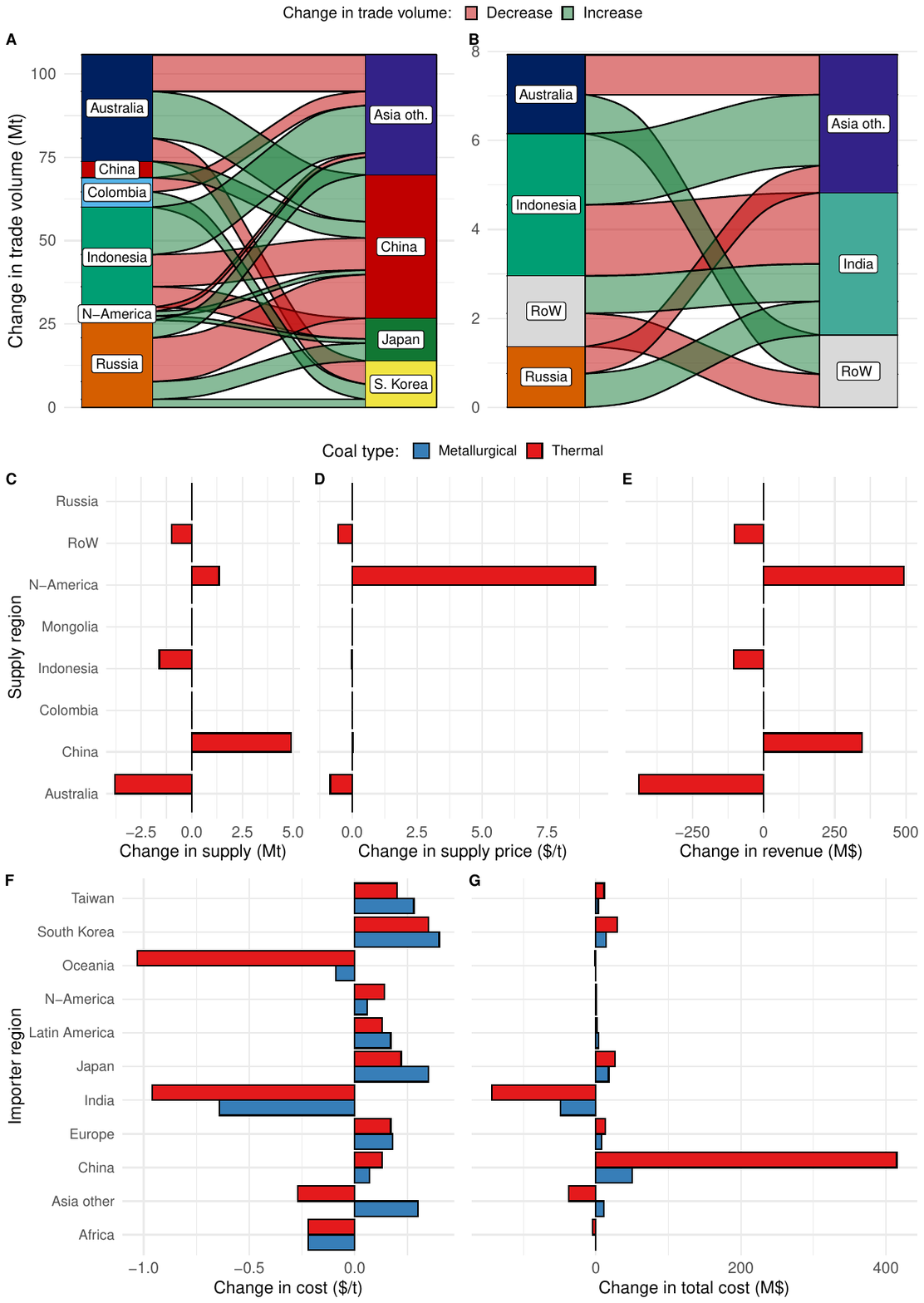}
        \caption{Changes to global coal trade in case of a restriction to maritime traffic through the Straits of Malacca and Indonesia. 
        \textmd{Changes in volumes of for thermal coal (A) and metallurgical coal (B) traded between key exporter and importer country (groups), supply volume (C), price (D) and revenue (E) for exporters, and cost per ton (F) and total costs (G) for importers. Note: trade flow changes $<$1 Mt in panel A and changes $<$0.5 Mt in panel B censored for legibility.}
        }
    \label{fig:malacca}
\end{figure}

Results indicate a small increase of thermal coal supplies from China to domestic consumers (Fig.~\ref{fig:malacca}A\&C). There is further re-organisation of supply chains with increasing exports from Russia's far east ports to Northern Asia, via routes which are unaffected by these closures, at the expense of supplies from Australia and Indonesia, which are on routes that would carry higher transport costs in this scenario. Effects on costs are very small, however, with cost increases at below 0.5 \$/t for all importing regions, and cost reductions of around 1 \$/t for a few smaller markets (Fig.~\ref{fig:malacca}F). Revenues are largely unchanged, apart from those for North American suppliers, who sit on the upper end of global supply curves, and whose exports are projected to increase by 1.35 Mt, at a price of about 9.5 \$/t higher than a scenario without restrictions to traffic through these Straits (Fig.~\ref{fig:malacca}C\&D).

\subsection{Taiwan Strait, South \& East China Sea restrictions}\label{taiwan-strait-south-east-china-sea-restrictions}

Restrictions to maritime traffic through the Taiwan Strait and South \& East China Sea would primarily affect China, the world's largest importer of coal. These waters transport between 300 and 500 Mtpa of seaborne imports headed for China over recent years\cite{SX2026Coal}, as well as several 100 Mt of coal which is loaded in China’s Northern ports and transported to coastal power and steel plants in China’s Southern provinces\cite{Gosens2022Chinas}.

Here we consider a scenario where international shipments through these waters would be restricted to 200 Mtpa, or half the 400 Mt of seaborne imports reported for 2025\cite{SX2026Coal}, whilst maritime traffic of Chinese ships along the coast may continue unrestricted.

\begin{figure}[H] % hbtp for here, top, bottom, next available page. H for really just here
    \centering
    \includegraphics[width=1.0\textwidth]{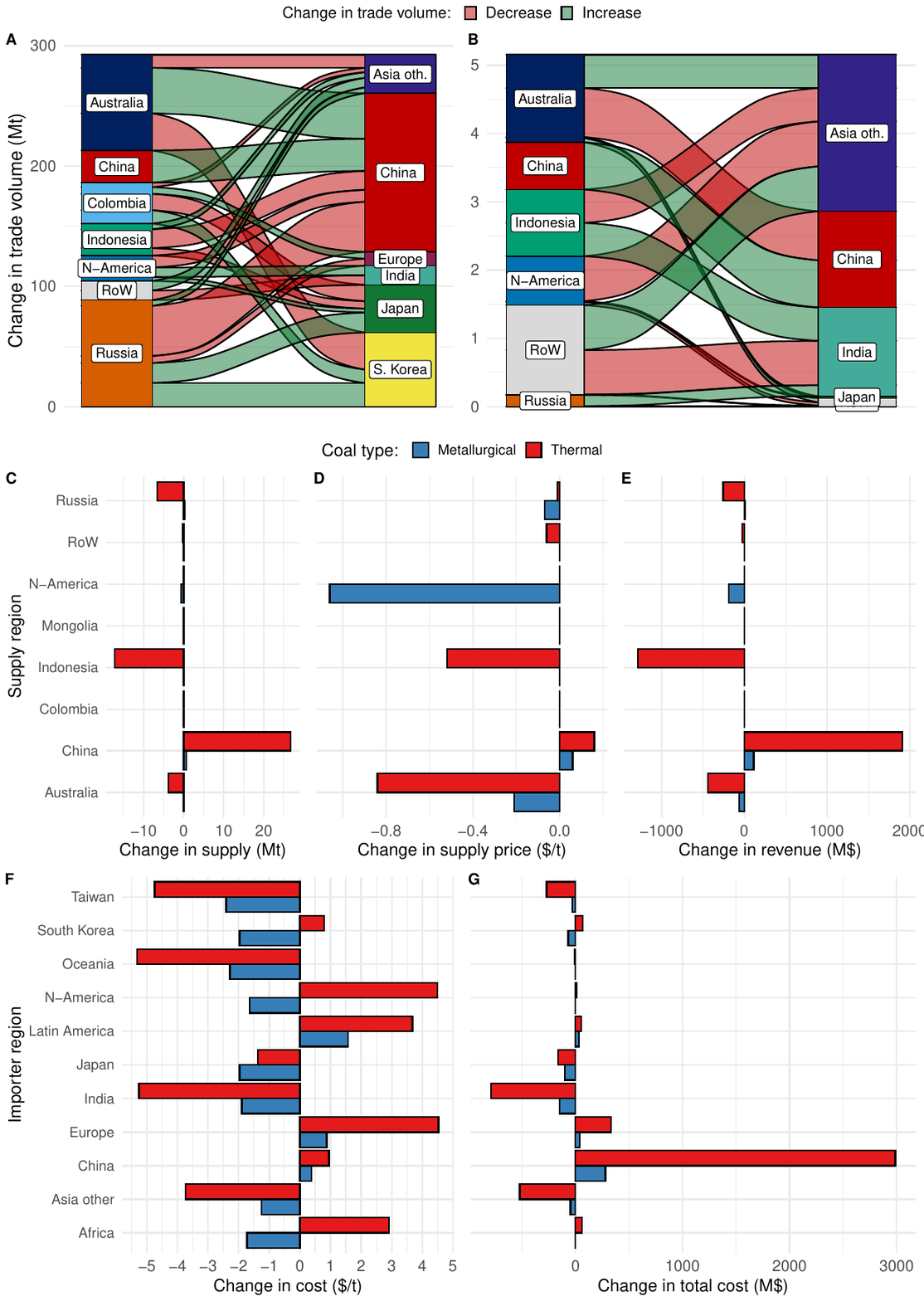}
        \caption{Changes to global coal trade in case of a restriction to maritime traffic through the South and East China Sea. 
        \textmd{Scenario for a restriction of Chinese seaborne imports to 200 Mt (approximately 50\% of baseline imports). Changes in volumes of for thermal coal (A) and metallurgical coal (B) traded between key exporter and importer country (groups), supply volume (C), price (D) and revenue (E) for exporters, and cost per ton (F) and total costs (G) for importers. Note: trade flow changes $<$6 Mt in panel A and changes $<$0.5 Mt in panel B censored for legibility.}
        }
    \label{fig:secs}
\end{figure}

In such a scenario, Chinese supplies of thermal coal ramp up by about 160 Mt, almost entirely at the expense of imports from Indonesia (Fig.~\ref{fig:secs}A\&C). Chinese supplies of metallurgical coal ramp up by about 4.5 Mt, mostly displacing Chinese imports from Australia (Fig. 5B). Subsequent re-organisation of Australian, Russian, and North American metallurgical coal flows, however, sees overall reduction in supplies from Australia limited to 1.5 Mt, whilst more expensive supplies from North America fall by about 3 Mt (Fig.~\ref{fig:secs}B\&C).

Despite the substantial ramp up of domestic supplies, marginal costs of thermal coal supplies to China rise only 4.5 \$/t, or about 5\%, and only about 2.5 \$/t or 1.7\% for metallurgical coal (Fig.~\ref{fig:secs}F), suggesting the Chinese coal supply and transport system is fairly resilient to international supply shocks. The removal of this Chinese demand for coal from global seaborne markets in turn generates more substantial cost reductions, of roughly between 5 to 10 \$/t of thermal and metallurgical coal, for most other consumers (Fig.~\ref{fig:secs}F). Revenue falls for most exporters, with a drop of almost 4 \$/t in Indonesian thermal coal prices leading to a reduction in revenue of a little over \$8 billion; whilst Australian and North American exporters also see substantial reductions in revenue (Fig.~\ref{fig:secs}D\&E).

\begin{figure}[H] % hbtp for here, top, bottom, next available page. H for really just here
    \centering
    \includegraphics[width=1.0\textwidth]{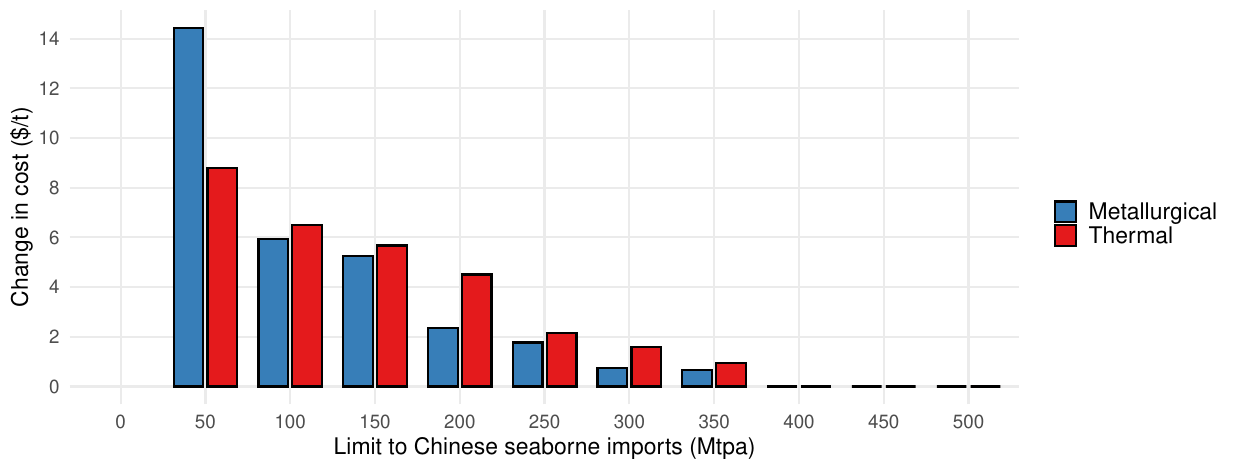}
        \caption{Change in marginal cost of thermal and metallurgical coal for Chinese consumers in case of restrictions to seaborne imports. 
        \textmd{A plot with changes in trade flows, and resulting revenue for exporters and costs for importers in a scenario where Chinese seaborne imports are limited to 50 Mt is provided in Supplementary Fig. \ref{fig:SECS50}. Note: no results could be provided for a scenario with 0 Mt seaborne imports as our data suggests that Chinese demand cannot be met from domestic coal mines and landborne imports alone.}
        }
    \label{fig:chn_restrictions}
\end{figure}

With further reductions to Chinese seaborne imports, costs increase for Chinese consumers ramp up further, to 9 \$/t for thermal and 15 \$/t for metallurgical coal for a scenario where seaborne imports are limited to 50 Mtpa (Fig.~\ref{fig:chn_restrictions}). We cannot report for results for a scenario where seaborne imports are completely cut off as our data suggests that Chinese demand cannot be met from domestic coal mines and landborne imports; in reality there will likely be further cost increases as Chinese domestic capacity would ramp up further at higher marginal costs.

In an alternative scenario, where despite restrictions to international maritime traffic in the South and East China Sea, ports in China's northern provinces would remain capable of receiving seaborne imports, cost increases for China are limited to less than 1 \$/t for thermal and less then 0.5 \$/t for metallurgical coal (Supplementary Fig. \ref{fig:SECSN}).

\section{Discussion}\label{discussion}
The global coal market is surprisingly resilient to potential disruptions from maritime chokepoints. Disruptions to traffic through the Panama or Suez Canal, or the Bosporus, Danish or Malacca Strait, result in increases of costs to importers of only 0.5 \$/t or less for both thermal and metallurgical coal. A number of smaller markets can see reductions in costs of up to several dollars per ton in some scenarios. The exception is a scenario where Chinese seaborne imports are severely restricted due to limits to traffic through the Taiwan Strait and South and East China seas. Costs of thermal coal are projected to increase by 4.5 \$/t, and metallurgical coal by about 2.5 \$/t, in the case of a halving of Chinese seaborne imports, which would simultaneously result in cost reductions of between 5 to 10 \$/t for both thermal and metallurgical coal, for most other importers. This still only represents a 5\% increase in thermal coal costs, and less than 2\% for metallurgical coal.

Changes in revenue for exporters are similarly limited to increases or decreases of roughly 0.5 \$/t or less, with two exceptions. First, when there is a genuine loss of access to markets, such as a restriction to Russian exports from its Black Sea ports, or reduced access to Chinese markets for Indonesian, North American, and Australian exporters, reductions in revenue can be around 5 \$/t. Second, North American exporters, which sit at the upper end of global supply curves, benefit from a number of scenarios for restricted trade flows, as they result in their revenues increasing by between 7 to 10 \$/t, though this concerns small total volumes of exports.

The effects on cost to importers and revenues for exporters are not uniform across the globe, as maritime chokepoints tend to create geographical separation of regional markets.

A first explanation for these limited effects is that none of the chokepoints analysed truly severs a substantial supply off the global seaborne market, as the closure of the Hormuz Strait does for gas and oil. Volumes traded through the Panama and Suez Canal, and the Bosporus and Danish Strait, are between circa 15 and 45 Mt in our baseline scenarios; or about 1-3\% of global seaborne coal
trade\cite{IEA2026CoalTrade}.

Trade through the Malacca Strait and alternative Straits between the islands of Indonesia is in the order of 100s of Mt, but the existence of several nearby alternative routes with limited additional transport costs limits impacts of the closure of either one of these Straits. Restrictions to seaborne imports by China, similarly 100s of Mt in usual years, can be replaced with domestic supplies at surprisingly limited additional cost.

Further moderating effects on costs and revenue is the re-routing of global coal trade flows, with exporters finding different markets and importers finding different suppliers, when maritime chokepoints close. This includes the effect that Russian rail connections, which allow some re-direction of supplies between markets on the Atlantic and in Asia.

Our analysis does not consider fuel switching. In power generation, short-term switching to alternative fuels like natural gas can be expected to be limited, however, unless excess generation capacity is available\cite{Entezari2026Assessment,Wilson2018Rapid}, and near impossible in steelmaking\cite{Wei2024Development}. Fuel switching may be further limited as prices for different fossil fuels are related\cite{Li2017Analysis,Villar2006relationship}, and chokepoints are likely to affect trade in several different fossil fuel markets simultaneously. More importantly, any fuel switching effects would very likely only lead to a further reduction in the limited effects on costs and revenues that we report here. Our analysis does not consider the possibility for changes in supply from domestic sources, in countries that both produce and import coal, apart from China; see the Methods section for more. That is, we assume imports in baseline and chokepoint scenarios are the same, again except for Chinese imports. The response in supply from domestic sources in these countries, however, would similarly only further moderate effects on costs and revenues that we report here.

The modeling approach used here could similarly produce insights on re-routing of trade flows, costs to importers, and revenues for exporters, in markets for gas, oil, and possibly other bulk commodities that are predominantly exported via seaborne.

Extensions of this model could be used to analyse cost-optimal levels of stockpiling, domestic supply capacity, and alternative transport capacity, when considering feasible supply shocks due to maritime chokepoints or other disruptions.

\section{Methods}\label{methods}
\subsection{Modelling approach}\label{modelling-approach}
In order to estimate flows of coal between different producers and consumers, we developed a source-sink transport model with a highly detailed representation of global transport networks for coal. This is an approach regularly applied to the modelling energy or resource markets\cite{Brown2018Synergies,Parzen2023PyPSA,Gosens2022Chinas,Paulus2011Coal,Goetschalckx2002Modeling,Leuthold2012Large,Kim2023Optimization}. The edges of this graph represent transport links such as navigation and rail networks, whilst the nodes represent producers (coal mines) and consumers of coal (power and steel plants primarily), as well as intermediate transport nodes such as ports or railway intersections. Provided that costs of production and transport are accurately estimated, such models are theorised to approximate real-world market outcomes\cite{Trutnevyte2016Does}.

Our model is a linear optimisation problem that minimises the cost of coal mining and transport to supply an exogenously determined level of demand for thermal and metallurgical coal, subject to constraints on mine production capacity, transport links capacity and port handling capacity, amongst others (see section `mathematical definition' below). The model is an extension of the Installation-Level China Coal Model\cite{Gosens2022installation} previously used to investigate changes to seaborne imports by the world biggest coal consumer\cite{Gosens2022Chinas}, with an extension to a full set of demand nodes representing other importers around the world, as well as a set of mines supplying the global seaborne market, and a representation of global seaborne routes now included. It is written in the Julia language, using the optimisation package JuMP\cite{Lubin2023JuMP}.

Our model reports on coal trade in time steps of a single year. Our baseline projections of trade flows and our scenarios with maritime chokepoints are all for the year 2026, with demand projected from historical (seaborne) import data, with growth projections from the IEA’s 2025 World Energy Outlook to extrapolate to 2026\cite{IEA2025World}. The reason for selecting this year is as follows. We aim to report on differences in global coal trade in the case of maritime chokepoints, versus a world where trade is uninterrupted. In the past few years, however, there have been political embargoes and other disruptions that have hindered the usual cost-optimising logic of the global coal market (more in the section on ‘Calibrating model outcomes to observed coal trade flows’ below). Rather than trying to replicate coal trade patterns observed in the real world in 2025, which would mean creating a model that accounts for these temporary disruptions of the usual market logic, we project results for a world free from these disruptions. The last year that can be argued to have been the case is the year 2019. We use this year for calibration of our model to observed real world trade, specifically to derive a set of preference parameters for trade between each exporter-importer country (group) pair. This set of preference parameters is then used as model input for projection years (explained in further detail in the section on ‘Calibrating model outcomes to observed coal trade flows’ below). 

\subsection{Model components and data sources}\label{model-components-and-data-sources}
\subsubsection{Network components}\label{network-components}
The location and production capacity of coal mines is sourced from Wood MacKenzie, one of several commercial suppliers of mining industry costing data. We use their data from the China coal supply data\cite{Wood2021China}, which covers mainland China and Mongolia, the Russia coal supply data\cite{Wood2024Russia}, and the seaborne export data sets for both thermal and metallurgical coal\cite{Wood2024GlobalMet,Wood2024GlobalTherm}.

Demand for coal is located in nodes that represent individual provinces of China, individual countries such as India, Japan, or South Korea, or country groups such as mainland EU or Southeast Asia, with grouping determined on the total volume of demand and geographic proximity; a full list of demand nodes, the countries they include, and the volume of imports they represent, is included in Supplementary Table \ref{tab:demandbynode}. A map of these demand nodes is provided in Supplementary Fig. \ref{fig:demand}.

The location and handling capacity of ports for exports of coal is taken from the Wood MacKenzie data sets\cite{Wood2024Russia,Wood2024GlobalMet,Wood2024GlobalTherm}. For coal terminals for imports, we source location and capacity data either from the Wood MacKenzie data set, for China\cite{Wood2021China}, or from Kpler\cite{Kpler2024Dry} for other countries, where we use the largest port by volume of imports in each country or country group as the location of the import node.

The navigation network between ports, and a small number of intermediary navigational waypoints, is created as a set of all origin-destination pairs, with distances calculated for each using on online sea route distance calculator\cite{SeaRoutesNavSea}. We also include a number of key railway networks. First, the entire Chinese railway network, with exact routes collected from the China Railway Map\cite{Peng2026China} and rail line capacities from Wood MacKenzie data\cite{Wood2021China}. We also include representations of key rail lines in Mongolia and Russia connecting coal mines in these countries to markets in China, the European Union, and ports in Russia's West and far East, using both routes and capacity data from Wood MacKenzie data\cite{Wood2024Russia,Wood2021China}. For the US, we include key rail lines, specifically in order to allow transport by rail to replace transit via the Panama Canal in our scenario where this Canal would be closed, using data on rail routes from the US department of Transportation\cite{US2026Maps}. A map of the rail network is provided in Supplementary Fig. \ref{fig:railnetwork}. For the Chinese transportation network, we also include the road networks and UHV transmission lines as included in the Installation-Level China Coal Model\cite{Gosens2022installation}, as well as the power and steel plants nodes in China, which connect with provincial level demand nodes, and largely have the function to very precisely represent the specific geographical locations where coal is consumed within China.

\subsubsection{Costing}\label{costing}
The costing data for coal mining for China, Mongolia, Russia, and mines supplying the seaborne market, is taken from the relevant data sets from Wood MacKenzie\cite{Wood2024Russia,Wood2021China,Wood2024GlobalMet,Wood2024GlobalTherm}. Different quality of both thermal and metallurgical coals fetch different prices, which we represent by subtracting these price premia from production costs. For thermal coal, we assume a price premium of US\$8/t for every 1,000 kcal/kg relative to the average energy content of 5,000 kcal/kg for a ton of coal traded in seaborne markets. For Hard Coking Coal (HCC), we assume price premia of up to \$32/t for coal with a Coke Strength after Reduction (CSR) number of 70 or higher, compared to low-grade HCC with a CSR of 55. For Semi-soft Coking Coal and PCI we do not presume any price premia. The values for these premia were derived from price and coal quality data from the Wood MacKenzie datasets\cite{Wood2024Russia,Wood2021China,Wood2024GlobalMet,Wood2024GlobalTherm}, with a methodology described in more detail in Gosens et al (2022)\cite{Gosens2022Chinas}.

The original cost data from Wood MacKenzie represents a cost curve with relatively large blocks of supply at the same cost, for individual mines. This is an issue when investigating switching between suppliers in an optimisation model such as ours, as there may be tipping points where a demand node may switch out all of its supply from one mine to another. We moderate this issue by adjusting costs for each individual mine by applying creating a small slope to the marginal supply costs of each mine, with a 5 per cent increase in production cost for the last versus the first ton of coal form each mine, whilst keeping the average production cost exactly as reported in the original data set.

Seaborne freight rates are based on dry bulk forward freight agreements reported by Simpson Spence Young\cite{Simpson2022Dry}, and are distinct for different routes, considering limitations to the size of ships that may traverse the Panama or Suez Canals\cite{Wilhemsen2026Panama,Wilhemsen2026Suez}. Port handling costs are from the Wood MacKenzie datasets\cite{Wood2024Russia,Wood2021China,Wood2024GlobalMet,Wood2024GlobalTherm}. Costs for rail transport are collected from different sources for China and Mongolia\cite{Rioux2016,NDRC2017Notice}, and Russia and the US\cite{Parming2023International}.

\subsubsection{Demand}\label{demand}
Demand for coal is located in the node and link network (or graph) in our model in a set of demand nodes that represent individual provinces of China, individual countries such as India, Japan, or South Korea, or country groups such as mainland EU or Southeast Asia, with grouping determined on the total volume of demand and geographic proximity (Supplementary Table \ref{tab:demandbynode}).

We set demand levels separately for thermal and metallurgical coal. We focus on seaborne imports, which represent 90\% of global coal trade\cite{IEA2026CoalTrade}. For the Chinese market, we represent total demand for coal, whether supplied from domestic sources, seaborne imports, or rail-borne imports from Mongolia and Russia. For the mainland EU region, we represent demand for seaborne imports plus rail-borne imports from Russia. For other markets we represent demand for seaborne imports only.

The choice to include Chinese domestic production and demand but not for other regions is justified as follows. Our analysis is of seaborne trade flows and how they change when maritime chokepoints appear. If supply is disrupted and prices rise in certain markets, countries could reduce imports and rely more strongly on domestic production of coal, for countries that have a domestic coal mining sector. The Chinese market represents 54\% of global coal consumption, and roughly 30\% of global imports\cite{IEA2026CoalInfo}. A number of further major coal consumers, e.g., Japan, Korea, Taiwan, cannot switch to domestic production as they have none; these countries represent circa a further 30\% of global imports. Seaborne imports by countries that also produce coal domestically, other than China, represent circa 40\% of global seaborne imports (Supplementary Table ~\ref{tab:importsthermal}\&~\ref{tab:importsmetcoal}). For countries in the latter group, we presume that rising prices in seaborne markets and a subsequent switching to domestic sources of coal will have a limited effect on their demand for seaborne imports. Rather, we presume that the effect on switching between different suppliers in the seaborne market in our chokepoint scenarios will dominate. This choice is also motivated by a trade-off between the expected limited additional insights and the excessive cost for separate datasets for domestic coal mining in each of these countries, on top of the ones for China, Mongolia, Russia, and mines supplying the seaborne market\cite{Wood2024Russia,Wood2021China,Wood2024GlobalMet,Wood2024GlobalTherm} that we use here. Further, our model does not consider fuel switching. In power generation, short-term switching to alternative fuels like natural gas can be expected to be limited, however, unless excess generation capacity is available\cite{Entezari2026Assessment,Wilson2018Rapid}, and near impossible in steelmaking\cite{Wei2024Development}.

We project demand for the year 2026. We use data on historical levels of demand and imports for China from SXCoal\cite{SX2026Coal} (a supplier of statistics on the Chinese coal industry), data on seaborne imports for other countries from Kpler\cite{Kpler2024Dry} (a service that provides statistics in trade data for energy and other products), and data on Russian imports by the mainland EU region from the IEA\cite{IEA2026CoalInfo}. Our dataset on seaborne coal trade from Kpler is current through mid 2023; we extrapolate by using growth in total imports from the IEA coal information dataset\cite{IEA2026CoalInfo}, for which we have the most recent version, which reports through to 2025; this is justified as seaborne imports are 90\% of total imports globally\cite{IEA2026CoalTrade}. We extrapolate 2026 demand from the last year of demand data available, with growth rates as projected in the IEA's 2025 World Energy Outlook, under the current policies scenario\cite{IEA2025World}.

Demand for thermal coal is determined in energy content (PJ) rather than Mt of coal, as different varieties of coal have different energy content, ranging between 3,000 and 7,000 kcal/kg in our mine level datasets of coal qualities\cite{Wood2024Russia,Wood2021China,Wood2024GlobalTherm}. For the Chinese market, we use historical demand data, expressed in energy units, from SX Coal\cite{SX2026Coal}, which splits demand for thermal coal into power generation and industrial uses. The power plant nodes in our network transform the thermal coal into electricity, with data on location, power plant capacity, and conversion efficiency from Global Energy Monitor\cite{Global2021Global}. For demand nodes in our global network outside of China we estimate the energy content of the seaborne imports, which are expressed in Mt only in the original trade data from Kpler\cite{Kpler2024Dry}. We do have precise data on calorific content for each of the coal mines in our dataset\cite{Wood2024Russia,Wood2021China,Wood2024GlobalTherm}, and coal qualities are very strongly linked to geology, with very different average calorific content by exporting region (either countries or provinces of e.g., Australia or Indonesia). We estimate the energy content of coal imports for each country by multiplying that country's volume of imports from each exporter region with the average energy content of coal mined in each of those exporter regions.

Demand for metallurgical coal in determined in Mt, but is split in three varieties, of Hard Coking Coal (HCC), Semi-Soft Coking Coal (SSCC) and Pulverised Coal for Injection (PCI). For the Chinese market, we use an average mix of 563 kg of HCC, 182 kg of SSCC, and 191 kg PCI for every ton of steel, as reported by Wood Mackenzie\cite{Wood2021China}. For demand nodes in our global network outside of China we estimate this mix in a manner equivalent to that used for the energy content of thermal coal imports. We estimate the mix of coking coal imports for each country by multiplying that country's volume of imports from each exporter region with the average mix of coking coal types mined in each of the exporter regions included in our model.

\subsection{Calibrating model outcomes to observed coal trade flows}\label{calibrating-model-outcomes-to-observed-coal-trade-flows}
The level of accuracy with which baseline predictions produced by our model reproduce observed coal trade patterns, determine in part the confidence we can place in the results it produces in scenarios that impose maritime chokepoints, including results on the re-routing of trade flows between exporters and importer pairs, and how it might affect costs to consumers and revenue and profits for producers.

We estimate this accuracy by comparing our model outcomes with observed real world data, in term of how well volumes of trade of both thermal and metallurgical coal between 7 exporter and 8 importer countries or country groups are replicated (Supplementary Fig. \ref{fig:calibration}). The model, before calibration, replicates total export volumes by exporter country (groups) relatively well, but is less precise in replicating country flows between individual exporter-importer pairs. We apply three calibration steps to improve this fit.

First, for a number of importer countries (or groups), the uncalibrated model tends to project very limited import diversification, with some countries importing very large shares, of up to 100\%, from individual suppliers. This is a common result of cost-minimization modelling of trade flows\cite{Haftendorn2010Modeling}. We therefore include a constraint that mimics import diversification strategies as previously considered in other models of coal or other energy trade\cite{Haftendorn2010Modeling}. Whilst import diversification is usefully measured in metrics such as the Herfindahl-Hirschman Index (HHI), we use a simpler constraint limiting the maximum share of imports from a single origin for each importer country to maximum observed historical shares, with these maximum shares determined separately for thermal and metallurgical coal. We believe it is more realistic, as well as conversative, to assume that national or corporate diversification strategies aim to limit dependency on single suppliers, rather than aspire a certain makeup of supplier shares.

Second, we consider that coal-fired power plants are designed or calibrated to operate on a specific quality of coal, in particular with regards to calorific content\cite{Gao2023Ultra}. To implement a constraint reflecting this, we use the information on the average calorific value of imports by each region, calculated as described in our section on Demand. To prevent making this constraint to restrictive, we allow a deviation from this optimal calorific value of plus or minus 500 kcal/kg in the blend of coals consumed.

These two calibration steps help improve the fit between observed and modelled exporter-importer pair trade flows, with the sum of squared errors (SSE) dropping from 30,994 to 22,043.

To further improve the fit, we apply a set of `preference parameters' for trade between different exporter-importer pairs, implemented as a reduction or increase to costs for a trade flow traversing a link between each exporter-importer pair. This is equivalent to Armington elasticities as used in CGE models of global trade\cite{Shiells1993Armington}, or country pair fixed effects in gravity models of trade\cite{Glick2016}; both of which capture unexplained differences in preferences for seemingly equivalent products from different trade partners. In some constrained optimisation problems such preference parameters (or shadow prices) could be derived by utilising the dual of our optimization problem, and changing the model variables on exporter-importer trade flows into constraints instead\cite{Romeijn1998Shadow}. We cannot utilise this functionality of dual optimisation problems in our case, however, as each exporter region can export to a list of importer regions. That means that a decrease in the costs applied to a link between one specific exporter-importer pair simultaneously affects the relative costs of exports to all other importer regions, which further propagates as this affects the relative  costs of imports for those importer regions from all exporter regions.

The only solution to such a problem is with iterative methods, or so-called swarm intelligence algorithms\cite{a2017Nature}. A range of such algorithms exist, but the underlying logic is the same; the algorithm attempts a number of randomly selected variations to the set of preference parameters, calculates the goodness of fit for each attempt, uses this information to identify solution regions with improved fit, and concentrates further attempts in this region. We select two commonly used algorithms, grey wolf optimisation and particle swarm optimisation\cite{Mirjalili2019Evolutionary}. We let these iterate until no further improvement to the goodness of fit are found over a 24-hour period. The application of these methods reduced the SSE between observed and modelled trade flows between all exporter-importer pairs to 3,314 (Supplementary Fig. \ref{fig:calibration}).

The resulting preference parameters are between -6.88 \$/t and 4.24 \$/t for thermal coal, with an average absolute value of 2.23 \$/t. For metallurgical coal they are between -3.22 \$/t and 3.42 \$/t, with an average absolute value of 1.63 \$/t. These are relatively small compared to the average cost of 87.8 \$/t for thermal coal and 147.4 \$/t for metallurgical coal in our baseline scenario, suggesting the model was well specified before application of these preference parameters.

We need to add here that a number of issues in global coal markets in recent years hamper the determination of these preference parameters, when the outcome is meant to reflect cost-optimised global trade. China has put an embargo on imports of Australian coal since 2020, and while unofficially lifted since, metallurgical coal exports to China have still far from recovered. A group of OECD countries has further imposed an embargo on imports of Russian energy, whilst implementation has not been entirely strict. The COVID-19 crisis has impacted both transport and mining sectors for several years. Lastly, China has further been increasing coal stockpiles by about 600 Mt over the last four years, in an effort to bolster energy security\cite{SX2026Coal}. Because of this, the last year in which global coal trade could be argued to follow mostly a cost-optimisation logic is 2019, and we use preference parameters derived for this year. These parameters correlate very well with parameters derived for the years 2017 and 2018 (Supplementary Fig. \ref{fig:prefparacorrelation}), but much less for those for later years, supporting the notion that the issues listed above were impacting the cost-optimising logic in global coal markets in these years.

\subsection{Determining costs and revenues}\label{determining-costs-and-revenues}
Our model is a linear optimisation problem that minimises total system cost, i.e., the sum of all coal mining and transport costs.

We derive the prices that consumers in importing regions would pay by first solving the optimisation problem with demand at baseline settings for all regions included in the model, and derive total system cost. After that we run the model again with demand reduced by 1 per cent for each of the importing regions in our model (listed in Supplementary Table \ref{tab:demandbynode}), separately for each region, and separately for both thermal and metallurgical coal, and again derive total system cost. We calculate the difference between the total system cost for each such run of the model and the baseline cost, and divide the difference between these two costs by the difference in the volume of coal consumed in both (i.e., resulting from the 1 per cent reduction in demand in the selected demand region). This results in the marginal cost of either thermal and metallurgical coal in each region. We repeat the exercise for the baseline scenario, and for scenarios with closure of chokepoints, to determine the difference in marginal costs for each coal type in each region.

For revenues, we consider that all producers fetch the same marginal cost, at any export port. Revenues are thus calculated as Free On Board (FOB), i.e., excluding any seaborne transport costs, which would not accrue to coal mines in the real world.

Costs and revenues for the three different types of metallurgical coal considered in this model (Hard Coking Coal, Semi-Soft Coking Coal, and Pulverised Coal for Injection) are calculated separately, before being summed as total costs or revenues for metallurgical coal in the results reported here.

\subsection{Scenario definition and selection}\label{scenario-definition-and-selection}
\begin{figure}[h] % hbtp for here, top, bottom, next available page. H for really just here
    \centering
    \includegraphics[width=1.0\textwidth]{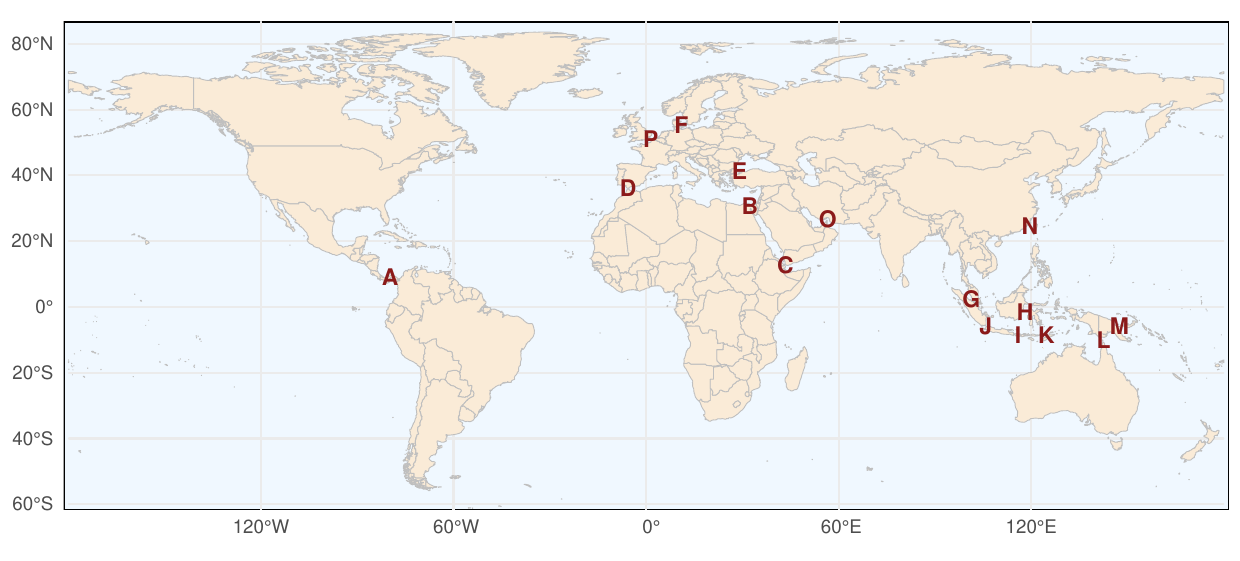}
        \caption{Map of global maritime chokepoints. 
        \textmd{Note: letters for each chokepoint correspond with those in Table ~\ref{tab:chokepoints}.}
        }
    \label{fig:map_chokepoints}
\end{figure}

We review the available literature on maritime chokepoints in energy trade\cite{UNCTAD2024Navigating,Pratson2023Assessing,Ulrichsen2026Maritime,Verschuur2025Systemic,Emmerson2012Maritime,Wang2024Assessing} and compile a list of all chokepoints mentioned (Fig.~\ref{fig:map_chokepoints}).

We then evaluate this list of chokepoints for relevance to the global coal trade. This depends on the volume of traffic through them, as well as how easily traffic may be rerouted\cite{Meza2026Implications,Wang2024Assessing,Bailey2017Chokepoints}. Bailey \& Wellesley\cite{Bailey2017Chokepoints} categorise maritime chokepoints as having a `very high' criticality when they are the sole route between two bodies of water, as `high' when alternative routes others require significantly longer detours when cut off, and as `moderate' when only minimal delays or costs are incurred from rerouting traffic. We classify each of the identified chokepoints accordingly and provide further explanation for their selection in this study (Table ~\ref{tab:chokepoints}).

\begin{table}
\caption{Description of maritime chokepoints and their relevance to global coal trade}
\label{tab:chokepoints}
\scriptsize
\centering
\begin{tblr}{
  width = \linewidth,
  colspec = {Q[230]Q[108]Q[76]Q[142]Q[383]},}
          Name & Connects & Criticality & Coal trade & Explanation for selection or exclusion \\ \hline
        Panama Canal (A) & Atlantic and Pacific Ocean & High & Circa 15 Mtpa\cite{Canal2023Principal} & Has already proven to be relevant chokepoint as climate-change induced reduction in rainfall and increased evaporation have reduced traffic \\ \hline
        Suez Canal (B), Bab Al-Mandab (C), and the Strait of Gibraltar (D) & Atlantic and Indian Oceans  & High & Between 15 and 45 Mtpa\cite{Kpler2024Update} & Has already proven to be relevant chokepoint with recent blockade and continuing threats from piracy and rebel groups. Considered together here as either would cut traffic between the Atlantic and Indian Oceans \\ \hline
        Bosporus Strait (or Turkish Strait; E) & Black Sea and Mediterranean Sea & High /Very high & Circa 25 Mtpa from Russian and Ukrainian ports on the Black Sea\cite{Kpler2024Update} & Credible chokepoint given recent military conflict in the Black Sea; alternative landborne routes exist via Russia’s Northern and Eastern ports, though rail and port capacity is limited \\ \hline
        Danish Straits (F) & Baltic Sea and North Sea, Atlantic Ocean & High/Very high & Circa 30 Mt from Russian Baltic ports\cite{SXCoal2026Russian} & Exports from ports in the Baltic (Ust-Luga, St Petersburg) could be rerouted to Russia’s Northern, Black Sea, and Eastern ports, though rail and port capacity is limited. There is no real credible scenario for a restriction to trade of this chokepoint, however \\ \hline
        Straits of Malacca and Singapore (G), Makassar Strait (H), Lombok strait (I), Sunda Strait (J), Ombai Strait (K), Torres Strait (L), and Vitiaz Strait (M) & Indian ocean and Oceania with South \& East China Sea & Moderate & Several 100 Mtpa between different countries in Asia and Oceania\cite{IEA2026CoalTrade,Kpler2024Dry} & There are several alternative routes close to each other, which are unlikely to be restricted simultaneously. Considered together here as either would result in detours for some coal trade between Indian Ocean and Oceania, and South \& East China Seas \\ \hline
        South China Sea, East China Sea, and the Taiwan Strait (N) & East \& South Asia & High & Several 100 Mtpa between different countries in Asia and Oceania and beyond\cite{IEA2026CoalTrade,Kpler2024Dry} & Credible chokepoint given complex geopolitical tensions between several nations in the region. Considered together here as either would result in disruption of seaborne imports by China \\ \hline
        Strait of Hormuz (O) & Persian Gulf \& Gulf of Oman & Very high & None & Excluded in this study. Critical chokepoint for oil and gas but not coal  \\ \hline
        Dover Strait (or English Channel; P) & North Sea and Atlantic Ocean & Moderate & Unclear & Excluded in this study. Traffic can easily be re-routed around the UK \& Ireland, and there is no credible scenario for a restriction to trade of this chokepoint \\ \hline
\end{tblr}
\end{table}

\newpage
\subsection{Model definition}\label{model-definition}

Below is the description and notation of the model objective and constraints. The model sets, variables, and parameters are described in Table \ref{tab:model_componenets}.

The model objective is to minimise all costs of production of coal, transport of coal (including preference parameters for country pairs), and transmission of electricity:

\begin{multline*}
Minimise\ \sum_{i,c,t}^{}{supply\_ Mt}_{i,c,t}\: \times \: {prod\_ cost}_{i,c,t} \times 1e6 \: + \\
\sum_{i,j,c,t}^{}{transp\_ Mt}_{i,j,tm,c,t} \times \left( {handling\_ cost}_{k,k^{'},t} + {pref\_ para}_{i,j,c} + {transp\_ cost}_{tm,t} \times {distance}_{i,j} \right) \times 1e6 \: + \\
\sum_{i,j,v,t}^{}{{transp\_ Mt}_{i,j,v,t} \times {CV}_{v} \times 1e9 \times kcal\_ to\_ PJ \times {conv\_ eff}_{i,j} \times {transm\_ cost}_{i,j,t}}
\end{multline*}

Constraint in coal mine production capacity:

\[0 \geq {Supply\_ Mt}_{i,c,t} \leq {prod\_ capa}_{i,c,t}\]

Constraint in mass flows (mass balance; outflows of any coal type cannot exceed supply plus inflows for each node):

\[{Supply\_ Mt}_{i,c,t} + \sum_{i,j,c,t}^{}{transp\_ Mt}_{j,i,c,t} \geq \sum_{i,j,c,t}^{}{transp\_ Mt}_{i,j,c,t}\]

Constraint in energy balances and energy demand (supply plus inflows of energy content must be at least equal to demand plus outflows for each node):

\begin{multline*}
{Supply\_ Mt}_{i,c,v,t} \times {CV}_{v} \times 1e9 \times kcal\_to\_PJ + \\
\sum_{i,j,c,v,t}^{}{{transp_{Mt}}_{j,i,c,v,t} \times {CV}_{v} \times 1e9 \times kcal_{to_{PJ}}} \geq {thermal\_demand}_{i,t}\:+ \\
\sum_{i,j,c,v,t}^{}{transp\_ Mt}_{i,j,c,v,t} \times {CV}_{v} \times 1e9 \times kcal\_ to\_ PJ
\end{multline*}

Constraint in average calorific value of thermal coal consumed in demand nodes (has to be within a defined range):

\[{minCVavg}_{i,t} \geq \sum_{i,j,c,v,t}^{}{{transp\_Mt}_{i,j,c,v,t} \times {CV}_{v}} \div \sum_{i,j,c,t}^{}{transp\_Mt}_{i,j,c,t} \leq {maxCVavg}_{i,t}\]

Constraint in metallurgical coal demand and mass balance (supply plus inflows of each type of coking coal must be at least equal to demand plus outflows for each node):

\[{Supply\_ Mt}_{i,c,t} + \sum_{i,j,c,t}^{}{transp\_ Mt}_{j,i,c,t} \geq {met\_ demand}_{i,c,t} + \sum_{i,j,c,t}^{}{transp\_ Mt}_{i,j,c,t}\]

Constraint in supply diversification (imports for thermal and metallurgical coal are subject to maximum shares):

\[\sum_{x,m,y}^{}{transp\_ Mt}_{x,m,y} \div \sum_{m,y}^{}{transp\_ Mt}_{m,y} \leq {max\_ market\_ share}_{m,y}\]

Constraint in transport capacity (flow over links cannot exceed capacity):

\[\sum_{i,j,c,t}^{}{transp\_ Mt}_{i,j,c,t} \leq {transp\_ capa}_{i,j,t}\]

Constraint in port capacity (flow out of ports cannot exceed handling capacity):

\[\sum_{p,j,c,t}^{}{transp\_ Mt}_{p,j,c,t} \leq {port\_ capa}_{p,t}\]

There are further a number of constraints that apply only to the Chinese network.

Constraint in transmission capacity (from power plant to demand node and between provincial-level demand nodes, over the UHV transmission network):

\[\sum_{i,j,c,v,t}^{}{transp\_ Mt}_{i,j,c,v,t} \times {CV}_{v} \times 1e9 \times kcal\_ to\_ PJ \times {conv\_ eff}_{i,j} \leq {elec\_ capa}_{i,j,t}\]

Constraint in specific metallurgical coal mix used in each steel plant node (average mix of HCC, SSCC, and PCI):

\[\sum_{p,j,c,t}^{}{transp\_ Mt}_{p,j,c,t} \times {HCC}_{c} \div 0.563 = \sum_{p,j,c,t}^{}{transp\_ Mt}_{p,j,c,t} \times {SSCC}_{c} \div 0.182\]

and

\[\sum_{p,j,c,t}^{}{transp\_ Mt}_{p,j,c,t} \times {HCC}_{c} \div 0.563 = \sum_{p,j,c,t}^{}{transp\_ Mt}_{p,j,c,t} \times {PCI}_{c} \div 0.191\]

Lastly, a constraint that applies only in the scenario where we limit seaborne imports to China:

\[\sum_{x,cp}^{}{transp\_ Mt}_{x,cp} \leq max\_ chn\_ seaborne\_ imp\]

\begin{table}
\caption{Model sets, variables, parameters}
\label{tab:model_componenets}
\scriptsize
\centering
\begin{tblr}{
  width = \linewidth,
  colspec = {Q[70]Q[50]Q[30]Q[170]},
  }
    \hline
        Notation & Type & Unit & Description \\ \hline
        N & Set & - & Set of all nodes \\ \hline
        i $\in$ N & Elements of a set & - & Nodes that are origin in a link \\ \hline
        j $\in$ N & Elements of a set & - & Nodes that are destination in a link \\ \hline
        t $\in$ T = \{2015,…,2026\} & Set & - & Set of all years \\ \hline
        k $\in$ K = \{mine, railway node, … port, demand node\} & Elements of a set & - & Set of all types of nodes \\ \hline
        y $\in$ Y = \{Thermal, Metalurgical\} & Set & - & Set of types of coal classified as either thermal or metallurgical \\ \hline
        c $\in$ C = \{Thermal, HCC, SSCC, PCI\} & Set & - & Set of types of coal classified as either thermal or type of metallurgical coal (HCC, SSCC, PCI) \\ \hline
        v $\in$ V = \{3000,3250,…,7000\} & Set & kcal/kg & Set of values of the energy content of thermal coal, binned with 250 kcal/kg steps \\ \hline
        p $\in$ P $\subseteq$ N & Set & - & Set of all nodes that are ports \\ \hline
        cp $\in$ CP $\subseteq$ P & Set & - & Set of all nodes that are Chinese ports \\ \hline
        s $\in$ S $\subseteq$ N & Set & - & Set of all nodes that are steel plants \\ \hline
        x $\in$ X $\subseteq$ N & Set & - & Set of all nodes that are within a certain exporter region \\ \hline
        m $\in$ M $\subseteq$ N & Set & - & Set of all nodes that are within a certain importer region \\ \hline
        tm $\in$ TM = \{Rail, truck, river barge, ocean-going ship\} & Set & - & Set of all possible transportation modes with specific transport rates \\ \hline
        supply\_Mt\textsubscript{i,c,t} & Variable & Mt & Production of coal type c at node i and year t \\ \hline
        transp\_Mt\textsubscript{i,j,c,t} & Variable & Mt & Transport of coal type c between nodes i and j in year t \\ \hline
        thermal\_demand\textsubscript{i,t} & Parameter & PJ & Demand for energy in thermal coal in node i and year t \\ \hline
        minCVavg\textsubscript{i,t} & Parameter & kcal/kg & Minimum calorific value of the blend of thermal coals consumed in each node \\ \hline
        maxCVavg\textsubscript{i,t} & Parameter & kcal/kg & Maximum calorific value of the blend of thermal coals consumed in each node \\ \hline
        met\_demand\textsubscript{i,c,t} & Parameter & Mt & Demand for metallurgical coal of type c in node i and year t \\ \hline
       max\_market\_share\textsubscript{y,m} & Parameter & - & Maximum share of imports derived from a single exporter for coal type y by importer region m \\ \hline
        max\_chn\_seaborne\_imp & Parameter & Mt & Maximum volume of Chinese seaborne imports \\ \hline
        CV\textsubscript{v} & Parameter & kcal/kg & Calorific value associated with thermal coal in bin v \\ \hline
        kcal\_to\_PJ & Parameter & - & Conversion factor for energy content, from kcal/kg to PJ/Mt \\ \hline
        HCC\textsubscript{c} & Parameter & - & Dummy variable indicating coal type is Hard Coking Coal \\ \hline
        SSCC\textsubscript{c} & Parameter & - & Dummy variable indicating coal type is Semi-Soft Coking Coal \\ \hline
        PCI\textsubscript{c} & Parameter & - & Dummy variable indicating coal type is Pulverised Coal for Injection \\ \hline
        prod\_cost\textsubscript{i,c,t} & Parameter & \$/t & Production cost of coal type c at node i and year t \\ \hline
        prod\_capa\textsubscript{i,c,t} & Parameter & Mt & Production capacity of coal type c at node i and year t \\ \hline
        transp\_capa\textsubscript{i,j,t} & Parameter & Mt & Transport capacity between nodes i and j in year t \\ \hline
        port\_capa\textsubscript{i,t} & Parameter & Mt & Handling capacity of port in year t \\ \hline
        elec\_capa\textsubscript{i,j,t} & Parameter & PJ & Electrical transmission capacity between nodes i and j in year t \\ \hline
        stpt\_capa\textsubscript{p,t} & Parameter & MT & Production capacity of steel plant p in year t \\ \hline
        conv\_eff\textsubscript{i,j} & Parameter & - & Efficiency of electrical conversion or transmission between nodes i and j \\ \hline
        distance\textsubscript{i,j} & Parameter & km & Distance between nodes i and j \\ \hline
        transp\_cost\textsubscript{m,t} & Parameter & \$/tkm & Variable cost for a transport mode m in year \\ \hline
        pref\_para\textsubscript{i,j,,c} & Parameter & \$/t & Preference parameter for transport of coal type c between nodes I and j in year t \\ \hline
        handling\_cost\textsubscript{k,k',t} & Parameter & \$/t & Fixed handling charge for transport between node of type k and type k’ in year t \\ \hline
        transm\_cost\textsubscript{i,j,t} & Parameter & \$/PJ & Variable cost for a transmission of electrical energy between nodes i and j in year t \\ \hline
\end{tblr}
\end{table}

\newpage
\section*{Data availability}\label{data_availability}
The data used for this analysis is available in a Zenodo repository\cite{gosens_global_2026}; for proprietary third party data, data structures have been preserved but replaced with placeholder values. All input data is publicly available at the sources cited in the paper.

\section*{Code availability}\label{code_availability}
The code used for this analysis is available in a Zenodo repository\cite{gosens_global_2026}.

\bibliography{refs}

@misc{IEA2026Strait,
	note = {[Online; accessed 2026-06-28]},
	author = {{IEA}},
	year = {2026},
	title = {Strait of {Hormuz} - {Factsheet}},
	url = {https://www.iea.org/about/oil-security-and-emergency-response/strait-of-hormuz},
	howpublished = {https://www.iea.org/about/oil-security-and-emergency-response/strait-of-hormuz},
}

@misc{Mishra2026How,
	note = {[Online; accessed 2026-06-28]},
	author = {Mishra, Vibhu},
	year = {2026},
	title = {How the {Hormuz} crisis keeps disrupting kitchens, ports and paychecks \textbar{} {UN} {News}},
	url = {https://news.un.org/en/story/2026/05/1167548},
	howpublished = {https://news.un.org/en/story/2026/05/1167548},
}

@article{Semieniuk2025Best,
	author = {Semieniuk, Gregor and Weber, Isabella M. and Weaver, Iain S. and Wasner, Evan and Braun, Benjamin and Holden, Philip B. and Salas, Pablo and Mercure, Jean-Francois and Edwards, Neil R.},
	journal = {Energy Research \& Social Science},
	doi = {10.1016/j.erss.2025.104221},
	issn = {2214-6296},
	year = {2025},
	pages = {104221},
	title = {Best of times, worst of times: record fossil-fuel profits, inflation and inequality},
	volume = {127},
}

@article{Neri2023Energy,
	note = {[Online; accessed 2026-07-10]},
	address = {Rochester, NY},
	author = {Neri, Stefano and Busetti, Fabio and Conflitti, Cristina and Corsello, Francesco and Delle Monache, Davide and Tagliabracci, Alex},
	doi = {10.2139/ssrn.4849009},
	year = {2023},
	publisher = {Social Science Research Network},
	title = {Energy {Price} {Shocks} and {Inflation} in the {Euro} {Area}},
	type = {SSRN {Scholarly} {Paper}},
	url = {https://papers.ssrn.com/abstract=4849009},
}

@techreport{UNCTAD2024Navigating,
	author = {{UNCTAD}},
	doi = {10.18356/27082822-114A},
	year = {2024},
	title = {Navigating {Troubled} {Waters}: Impact to {Global} {Trade} of {Disruption} of {Shipping} {Routes} in the {Red} {Sea}, {Black} {Sea} and {Panama} {Canal}: UNCTAD {Rapid} {Assessment}},
	type = {United {Nations} {Conference} on {Trade} and {Development} ({UNCTAD}) {Policy} {Briefs}},
	url = {https://www.un-ilibrary.org/content/papers/10.18356/27082822-114A},
	howpublished = {https://www.un-ilibrary.org/content/papers/10.18356/27082822-114A},
}

@article{Pratson2023Assessing,
	author = {Pratson, Lincoln F.},
	journal = {Communications in Transportation Research},
	doi = {10.1016/j.commtr.2022.100083},
	issn = {2772-4247},
	year = {2023},
	month = {dec 1},
	pages = {100083},
	title = {Assessing impacts to maritime shipping from marine chokepoint closures},
	volume = {3},
}

@misc{S2023Panama,
	note = {[Online; accessed 2026-05-31]},
	author = {{S\&P Global Energy}},
	year = {2023},
	title = {Panama {Canal} transit reduction impacts thermal coal shipments in {Atlantic}: sources},
	url = {https://www.spglobal.com/energy/en/news-research/latest-news/coal/080223-panama-canal-transit-reduction-impacts-thermal-coal-shipments-in-atlantic-sources},
	howpublished = {https://www.spglobal.com/energy/en/news-research/latest-news/coal/080223-panama-canal-transit-reduction-impacts-thermal-coal-shipments-in-atlantic-sources},
}

@article{Ulrichsen2026Maritime,
	author = {Ulrichsen, Kristian Coates and Krane, Jim},
	year = {2026},
	title = {Maritime {Chokepoints} and {Risks} to {Global} {Shipping} and {Energy} {Security}. {Rice} {University}\textquoteright{}s {Baker} {Institute} for {Public} {Policy}, {March} 16, 2026.},
}

@article{Verschuur2025Systemic,
	author = {Verschuur, Jasper and Lumma, Johannes and Hall, Jim W.},
	journal = {Nature Communications},
	doi = {10.1038/s41467-025-65403-w},
	issn = {2041-1723},
	number = {1},
	year = {2025},
	pages = {10421},
	publisher = {Nature Publishing Group},
	title = {Systemic impacts of disruptions at maritime chokepoints},
	volume = {16},
}

@article{Brown2018Synergies,
	author = {Brown, T. and Schlachtberger, D. and Kies, A. and Schramm, S. and Greiner, M.},
	journal = {Energy},
	doi = {10.1016/j.energy.2018.06.222},
	issn = {0360-5442},
	year = {2018},
	pages = {720--739},
	title = {Synergies of sector coupling and transmission reinforcement in a cost-optimised, highly renewable {European} energy system},
	volume = {160},
}

@article{Parzen2023PyPSA,
	author = {Parzen, Maximilian and Abdel-Khalek, Hazem and Fedotova, Ekaterina and Mahmood, Matin and Frysztacki, Martha Maria and Hampp, Johannes and Franken, Lukas and Schumm, Leon and Neumann, Fabian and Poli, Davide and Kiprakis, Aristides and Fioriti, Davide},
	journal = {Applied Energy},
	doi = {10.1016/j.apenergy.2023.121096},
	issn = {0306-2619},
	year = {2023},
	pages = {121096},
	title = {PyPSA-{Earth}. {A} new global open energy system optimization model demonstrated in {Africa}},
	volume = {341},
}

@article{Gosens2022Chinas,
	author = {Gosens, Jorrit and Turnbull, Alex B.H. and Jotzo, Frank},
	journal = {Joule},
	doi = {10.1016/J.JOULE.2022.03.008},
	issn = {2542-4351},
	number = {4},
	year = {2022},
	pages = {782--815},
	publisher = {Cell Press},
	title = {China\textquoteright{}s decarbonization and energy security plans will reduce seaborne coal imports: Results from an installation-level model},
	volume = {6},
}

@article{Paulus2011Coal,
	author = {Paulus, Moritz and Tr{\" u}by, Johannes},
	journal = {Energy Economics},
	issn = {0140-9883},
	number = {6},
	year = {2011},
	pages = {1127--1137},
	publisher = {Elsevier},
	title = {Coal lumps vs. electrons: How do {Chinese} bulk energy transport decisions affect the global steam coal market?},
	volume = {33},
}

@techreport{Komiss2011Economic,
	author = {Komiss, William and Huntzinger, LaVar},
	year = {2011},
	title = {The {Economic} {Implications} of {Disruptions} to {Maritime} {Oil} {Chokepoints}},
}

@article{Meza2026Implications,
	author = {Meza, Abel and Ari, Ibrahim and Al Sada, Mohammed and Ko{\c c}, Muammer},
	journal = {Journal of Transportation Security},
	doi = {10.1007/s12198-026-00350-1},
	issn = {1938-775X},
	number = {1},
	year = {2026},
	pages = {28},
	title = {Implications of interrupting the {Hormuz} {Strait} in the {LNG} trade},
	volume = {19},
}

@techreport{Canal2023Principal,
	author = {{Canal de Panama}},
	year = {2023},
	title = {Principal {Commodities} {Shipped} {Through} the {Panama} {Canal} {Fiscal} {Years} 2023-2021},
	url = {https://pancanal.com/wp-content/uploads/2023/11/06-Principal-Commodities-Shipped-Through-the-Panama-Canal.pdf},
	howpublished = {https://pancanal.com/wp-content/uploads/2023/11/06-Principal-Commodities-Shipped-Through-the-Panama-Canal.pdf},
}

@techreport{Canal2026Estadisticas,
	author = {{Canal de Panama}},
	year = {2026},
	title = {Estad{\' i}sticas de {Tr}{\' a}nsito},
	url = {https://pancanal.com/estadisticas/},
	howpublished = {https://pancanal.com/estadisticas/},
}

@misc{Council2023Whats,
	note = {[Online; accessed 2026-05-31]},
	author = {{Council on Foreign Relations}},
	year = {2023},
	title = {What\textquoteright{}s {Causing} the {Panama} {Canal} {Logjam}},
	url = {https://www.cfr.org/articles/whats-causing-panama-canal-logjam},
	howpublished = {https://www.cfr.org/articles/whats-causing-panama-canal-logjam},
}

@misc{Kpler2024Update,
	note = {[Online; accessed 2026-06-20]},
	author = {{Kpler}},
	year = {2024},
	title = {Update on {Red} {Sea} trade flow impacts},
	url = {https://www.kpler.com/blog/update-on-red-sea-trade-flow-impacts},
	howpublished = {https://www.kpler.com/blog/update-on-red-sea-trade-flow-impacts},
}

@techreport{Wood2024Russia,
	author = {{Wood Mackenzie}},
	year = {2024},
	title = {Russia coal supply data - version of {Q2} 2024},
	url = {https://www.woodmac.com/industry/metals-and-mining/coal-product-suite/},
	howpublished = {https://www.woodmac.com/industry/metals-and-mining/coal-product-suite/},
}

@misc{EIA2026Strait,
	note = {[Online; accessed 2026-06-28]},
	author = {{EIA}},
	year = {2026},
	title = {The {Strait} of {Malacca}, a key oil trade chokepoint, links the {Indian} and {Pacific} {Oceans} - {U}.{S}. {Energy} {Information} {Administration} ({EIA})},
	url = {https://www.eia.gov/todayinenergy/detail.php?id=32452},
	howpublished = {https://www.eia.gov/todayinenergy/detail.php?id=32452},
}

@article{Yin2021Energy,
	author = {Yin, Yuwei and Lam, Jasmine Siu Lee},
	journal = {Maritime Business Review},
	doi = {10.1108/MABR-12-2020-0070},
	issn = {2397-3757},
	number = {2},
	year = {2021},
	pages = {145--160},
	title = {Energy strategies of {China} and their impacts on energy shipping import through the {Straits} of {Malacca} and {Singapore}},
	volume = {7},
}

@techreport{SX2026Coal,
	author = {{SX Coal}},
	year = {2026},
	title = {Coal {Import} \& {Export} {Data}},
	url = {https://www.sxcoal.com/en/data/package/FW1907B?isSubCode=FW6002D},
	howpublished = {https://www.sxcoal.com/en/data/package/FW1907B?isSubCode=FW6002D},
}

@techreport{IEA2026CoalTrade,
	note = {[Online; accessed 2026-07-08]},
	author = {{IEA}},
	year = {2026},
	title = {Coal 2025 {Analysis} and forecast to 2030},
	url = {https://www.iea.org/reports/coal-2025/trade},
	howpublished = {https://www.iea.org/reports/coal-2025/trade},
}

@article{Entezari2026Assessment,
	author = {Entezari, Negin and Fuinhas, Jos{\' e} Alberto},
	journal = {Energy Sources, Part B: Economics, Planning, and Policy},
	doi = {10.1080/15567249.2026.2652403},
	issn = {1556-7249},
	number = {1},
	year = {2026},
	note = {\textunderscore{}eprint: https://doi.org/10.1080/15567249.2026.2652403},
	pages = {2652403},
	publisher = {Taylor \& Francis},
	title = {Assessment of the impact of supply and demand shocks on sustained price increases: Evidence from the wholesale electricity market in {Portugal}},
	volume = {21},
}

@article{Wilson2018Rapid,
	author = {Wilson, I. A. Grant and Staffell, Iain},
	journal = {Nature Energy},
	doi = {10.1038/s41560-018-0109-0},
	issn = {2058-7546},
	number = {5},
	year = {2018},
	pages = {365--372},
	publisher = {Nature Publishing Group},
	title = {Rapid fuel switching from coal to natural gas through effective carbon pricing},
	volume = {3},
}

@article{Wei2024Development,
	author = {Wei, Chengzhi and Zhang, Xin and Zhang, Jin and Xu, Liangping and Li, Guanghui and Jiang, Tao},
	journal = {Engineering},
	doi = {10.1016/j.eng.2024.04.025},
	issn = {2095-8099},
	year = {2024},
	pages = {93--109},
	title = {Development of {Direct} {Reduced} {Iron} in {China}: Challenges and {Pathways}},
	volume = {41},
}

@article{Li2017Analysis,
	author = {Li, Hong and Chen, Long and Wang, Di and Zhang, Huize},
	series = {Proceedings of the 9th {International} {Conference} on {Applied} {Energy}},
	journal = {Energy Procedia},
	doi = {10.1016/j.egypro.2017.12.376},
	issn = {1876-6102},
	year = {2017},
	pages = {3141--3146},
	title = {Analysis of the {Price} {Correlation} between the {International} {Natural} {Gas} and {Coal}},
	volume = {142},
}

@article{Villar2006relationship,
	author = {Villar, Jose A. and Joutz, Frederick L.},
	journal = {Energy information administration, office of oil and gas},
	year = {2006},
	pages = {1--43},
	title = {The relationship between crude oil and natural gas prices},
	volume = {1},
}

@article{Goetschalckx2002Modeling,
	author = {Goetschalckx, Marc and Vidal, Carlos J. and Dogan, Koray},
	journal = {European Journal of Operational Research},
	doi = {10.1016/S0377-2217(02)00142-X},
	issn = {0377-2217},
	number = {1},
	year = {2002},
	pages = {1--18},
	title = {Modeling and design of global logistics systems: A review of integrated strategic and tactical models and design algorithms},
	volume = {143},
}

@article{Leuthold2012Large,
	author = {Leuthold, Florian U. and Weigt, Hannes and von Hirschhausen, Christian},
	journal = {Networks and Spatial Economics},
	doi = {10.1007/s11067-010-9148-1},
	issn = {1572-9427},
	number = {1},
	year = {2012},
	pages = {75--107},
	title = {A {Large}-{Scale} {Spatial} {Optimization} {Model} of the {European} {Electricity} {Market}},
	volume = {12},
}

@article{Kim2023Optimization,
	author = {Kim, Jin-Kuk and Park, Haryn and Kim, Se-jung and Lee, Joohwa and Song, Yongjae and Yi, Sung Chul},
	journal = {Renewable and Sustainable Energy Reviews},
	doi = {10.1016/j.rser.2023.113429},
	issn = {1364-0321},
	year = {2023},
	pages = {113429},
	title = {Optimization models for the cost-effective design and operation of renewable-integrated energy systems},
	volume = {183},
}

@article{Trutnevyte2016Does,
	author = {Trutnevyte, Evelina},
	journal = {Energy},
	doi = {10.1016/j.energy.2016.03.038},
	issn = {0360-5442},
	year = {2016},
	pages = {182--193},
	title = {Does cost optimization approximate the real-world energy transition?},
	volume = {106},
}

@article{Gosens2022installation,
	title = {A protocol to determine the least cost supply of coal to {China} with an installation-level optimization model},
	volume = {3},
	issn = {2666-1667},
	url = {https://www.cell.com/star-protocols/abstract/S2666-1667(22)00753-5},
	doi = {10.1016/j.xpro.2022.101873},
	language = {English},
	number = {4},
	urldate = {2026-07-31},
	journal = {STAR Protocols},
	publisher = {Elsevier},
	author = {Gosens, Jorrit and Turnbull, Alex B. H.},
	year = {2022},
}

@article{Lubin2023JuMP,
	author = {Lubin, Miles and Dowson, Oscar and Garcia, Joaquim Dias and Huchette, Joey and Legat, Beno{\^ i}t and Vielma, Juan Pablo},
	journal = {Mathematical Programming Computation},
	doi = {10.1007/s12532-023-00239-3},
	issn = {1867-2957},
	number = {3},
	year = {2023},
	pages = {581--589},
	title = {JuMP 1.0: recent improvements to a modeling language for mathematical optimization},
	volume = {15},
}

@techreport{Wood2021China,
	author = {{Wood Mackenzie}},
	year = {2021},
	title = {China coal supply data - version of {Q1} 2021},
	url = {https://www.woodmac.com/industry/metals-and-mining/coal-product-suite/},
	howpublished = {https://www.woodmac.com/industry/metals-and-mining/coal-product-suite/},
}

@techreport{Wood2024GlobalMet,
	author = {{Wood Mackenzie}},
	year = {2024},
	title = {Global seaborne export metallurgical coal cost curve},
	url = {https://www.woodmac.com/reports/coal-global-seaborne-export-metallurgical-coal-cost-curve-37910646},
	howpublished = {https://www.woodmac.com/reports/coal-global-seaborne-export-metallurgical-coal-cost-curve-37910646},
}

@techreport{Wood2024GlobalTherm,
	author = {{Wood Mackenzie}},
	year = {2024},
	title = {Global seaborne export thermal coal cost curve},
	url = {https://www.woodmac.com/reports/coal-global-seaborne-export-thermal-coal-cost-curve-37910732},
	howpublished = {https://www.woodmac.com/reports/coal-global-seaborne-export-thermal-coal-cost-curve-37910732},
}

@techreport{Kpler2024Dry,
	author = {{Kpler}},
	year = {2024},
	title = {Dry bulk predictive analytics \& insights},
	url = {https://www.kpler.com/product/commodities/dry-bulk-flows-and-insight},
	howpublished = {https://www.kpler.com/product/commodities/dry-bulk-flows-and-insight},
}

@misc{SeaRoutesNavSea,
	note = {[Online; accessed 2026-07-12]},
	author = {{SeaRoutesNav}},
	title = {Sea {Route} {Calculator} -- {Port} {Distance} \& {Voyage} {Time}},
	url = {https://searoutesnav.com/},
	howpublished = {https://searoutesnav.com/},
}

@misc{Peng2026China,
	note = {[Online; accessed 2026-07-12]},
	author = {Peng, Kejing},
	year = {2026},
	title = {China {Railway} {Map}},
	url = {http://cnrail.geogv.org/enus/about},
	howpublished = {http://cnrail.geogv.org/enus/about},
}

@misc{US2026Maps,
	note = {[Online; accessed 2026-07-12]},
	author = {{US Department of Transportation}},
	year = {2026},
	title = {Maps -- {Geographic} {Information} {System} \textbar{} {FRA}},
	url = {https://railroads.dot.gov/rail-network-development/maps-and-data/maps-geographic-information-system/maps-geographic},
	howpublished = {https://railroads.dot.gov/rail-network-development/maps-and-data/maps-geographic-information-system/maps-geographic},
}

@techreport{Simpson2022Dry,
	author = {{Simpson Spence Young}},
	year = {2022},
	title = {Dry bulk forward freight agreements},
}

@techreport{Wilhemsen2026Panama,
	author = {{Wilhemsen}},
	year = {2026},
	title = {Panama {Toll} {Calculator}},
	url = {https://www.wilhelmsen.com/tollcalculators/panama-toll-calculator/CalculatePanama},
	howpublished = {https://www.wilhelmsen.com/tollcalculators/panama-toll-calculator/CalculatePanama},
}

@techreport{Wilhemsen2026Suez,
	author = {{Wilhemsen}},
	year = {2026},
	title = {Suez {Toll} {Calculator}},
	url = {https://www.wilhelmsen.com/tollcalculators/suez-toll-calculator/},
	howpublished = {https://www.wilhelmsen.com/tollcalculators/suez-toll-calculator/},
}

@article{Rioux2016,
	author = {Rioux, Bertrand and Galkin, Philipp and Murphy, Frederic and Pierru, Axel},
	journal = {Energy Economics},
	doi = {https://doi.org/10.1016/j.eneco.2016.10.013},
	issn = {0140-9883},
	year = {2016},
	pages = {387--399},
	title = {Economic impacts of debottlenecking congestion in the {Chinese} coal supply chain},
	volume = {60},
}

@misc{NDRC2017Notice,
	note = {[Online; accessed 2021-05-18]},
	author = {{NDRC}},
	year = {2017},
	title = {Notice of the {NDRC} on deepening the market-oriented reform of railway freight transport charges and other related issues. {NDRC} {Price} [2017] {No}. 2163},
	url = {https://www.ndrc.gov.cn/xxgk/zcfb/tz/201712/t20171226_962626.html},
	howpublished = {https://www.ndrc.gov.cn/xxgk/zcfb/tz/201712/t20171226\textunderscore{}962626.html},
}

@techreport{Parming2023International,
	author = {Parming, Veiko},
	year = {2023},
	title = {International {Comparison} of {Railway} {Freight} {Rates}},
	url = {https://www.railcan.ca/wp-content/uploads/2023/02/International-Comparison-of-Railway-Freight-Rates-2.pdf},
	howpublished = {https://www.railcan.ca/wp-content/uploads/2023/02/International-Comparison-of-Railway-Freight-Rates-2.pdf},
}

@techreport{IEA2026CoalInfo,
	author = {{IEA/OECD}},
	year = {2026},
	title = {Coal {Information}},
	url = {https://www.iea.org/data-and-statistics/data-product/coal-information-2},
	howpublished = {https://www.iea.org/data-and-statistics/data-product/coal-information-2},
}

@techreport{IEA2025World,
	author = {{IEA}},
	year = {2025},
	note = {collection-title: World Energy Outlook},
	pages = {1--519},
	institution = {International Energy Agency},
	title = {World {Energy} {Outlook} 2025},
	url = {iea.li/WEO2025},
}

@misc{Global2021Global,
	note = {[Online; accessed 2021-06-22]},
	author = {{Global Energy Monitor}},
	year = {2021},
	title = {Global {Coal} {Plant} {Tracker}},
	url = {https://endcoal.org/tracker/},
	howpublished = {https://endcoal.org/tracker/},
}

@article{Haftendorn2010Modeling,
	author = {Haftendorn, Clemens and Holz, Franziska},
	journal = {The Energy Journal},
	doi = {10.5547/ISSN0195-6574-EJ-Vol31-No4-10},
	issn = {0195-6574},
	number = {4},
	year = {2010},
	pages = {205--230},
	publisher = {SAGE Publications},
	title = {Modeling and {Analysis} of the {International} {Steam} {Coal} {Trade}},
	volume = {31},
}

@article{Gao2023Ultra,
	author = {Gao, Rui and Li, Jiaxuan and Wang, Shuqing and Zhang, Yan and Zhang, Lei and Ye, Zefu and Zhu, Zhujun and Yin, Wangbao and Jia, Suotang},
	journal = {Analytical Methods},
	doi = {10.1039/d2ay02086f},
	issn = {1759-9660},
	number = {13},
	year = {2023},
	pages = {1674--1680},
	title = {Ultra-repeatability measurement of calorific value of coal by {NIRS}-{XRF}},
	volume = {15},
}

@article{Shiells1993Armington,
	author = {Shiells, Clinton R. and Reinert, Kenneth A.},
	journal = {The Canadian Journal of Economics / Revue canadienne d'Economique},
	doi = {10.2307/135909},
	issn = {0008-4085},
	number = {2},
	year = {1993},
	pages = {299--316},
	publisher = {[Wiley, Canadian Economics Association]},
	title = {Armington {Models} and {Terms}-of-{Trade} {Effects}: Some {Econometric} {Evidence} for {North} {America}},
	volume = {26},
}

@article{Glick2016,
	author = {Glick, Reuven and Rose, Andrew K},
	journal = {European Economic Review},
	issn = {0014-2921},
	year = {2016},
	pages = {78--91},
	publisher = {Elsevier},
	title = {Currency unions and trade: A post-{EMU} reassessment},
	volume = {87},
}

@article{Romeijn1998Shadow,
	author = {Romeijn, H. Edwin and Smith, Robert L.},
	journal = {Mathematics of Operations Research},
	issn = {0364-765X},
	number = {1},
	year = {1998},
	pages = {239--256},
	publisher = {INFORMS},
	title = {Shadow {Prices} in {Infinite}-{Dimensional} {Linear} {Programming}},
	volume = {23},
}

@book{a2017Nature,
	note = {[Online; accessed 2026-07-26]},
	address = {Cham},
	series = {Modeling and {Optimization} in {Science} and {Technologies}},
	doi = {10.1007/978-3-319-50920-4},
	editor = {Patnaik, Srikanta and Yang, Xin-She and Nakamatsu, Kazumi},
	isbn = {978-3-319-50919-8},
	year = {2017},
	publisher = {Springer International Publishing},
	title = {Nature-{Inspired} {Computing} and {Optimization}: Theory and {Applications}},
	url = {http://link.springer.com/10.1007/978-3-319-50920-4},
	volume = {10},
}

@book{Mirjalili2019Evolutionary,
	note = {[Online; accessed 2026-07-26]},
	address = {Cham},
	author = {Mirjalili, Seyedali},
	series = {Studies in {Computational} {Intelligence}},
	doi = {10.1007/978-3-319-93025-1},
	isbn = {978-3-319-93024-4},
	year = {2019},
	publisher = {Springer International Publishing},
	title = {Evolutionary {Algorithms} and {Neural} {Networks}},
	url = {http://link.springer.com/10.1007/978-3-319-93025-1},
	volume = {780},
}

@article{Emmerson2012Maritime,
	author = {Emmerson, Charles and Stevens, Paul},
	year = {2012},
	title = {Maritime {Choke} {Points} and the {Global} {Energy} {System}},
}

@article{Wang2024Assessing,
	author = {Wang, Xue and Du, Debin and Peng, Yan},
	journal = {Sustainability},
	doi = {10.3390/su16010384},
	issn = {2071-1050},
	number = {1},
	year = {2024},
	pages = {384},
	publisher = {Multidisciplinary Digital Publishing Institute},
	title = {Assessing the {Importance} of the {Marine} {Chokepoint}: Evidence from {Tracking} the {Global} {Marine} {Traffic}},
	volume = {16},
}

@book{Bailey2017Chokepoints,
	address = {London},
	author = {Bailey, Rob and Wellesley, Laura},
	series = {Chatham {House} report},
	isbn = {978-1-78413-230-9},
	year = {2017},
	publisher = {Chatham House},
	title = {Chokepoints and vulnerabilities in global food trade},
}

@misc{SXCoal2026Russian,
	note = {[Online; accessed 2026-07-09]},
	author = {{SXCoal}},
	year = {2026},
	title = {Russian thermal coal exports rise by 9\% {YoY} in 2025 - {Sxcoal}},
	url = {https://www.sxcoal.com/en/news/detail/2011718914165735426},
	howpublished = {https://www.sxcoal.com/en/news/detail/2011718914165735426},
}

@misc{gosens_global_2026,
	title = {Global coal trade and maritime chokepoints},
	url = {https://zenodo.org/records/21787231},
	doi = {10.5281/zenodo.21787231},
	urldate = {2026-08-04},
	publisher = {Zenodo},
	author = {Gosens, Jorrit and Turnbull, Alex and Jotzo, Frank},
	year = {2026},
    howpublished = {https://zenodo.org/records/21787231},
}

\newpage
\section*{Supplementary material}\label{Supplementary}
% Reset the figure counter to 0
\setcounter{figure}{0}
\setcounter{table}{0}
% Change "Figure" label to "Supplementary Figure"
\renewcommand{\figurename}{Supplementary Figure} 
\renewcommand{\tablename}{Supplementary Table} 

\begin{table}[!ht]
    \caption{Trade volumes, costs to consumers, and revenue, for different types of metallurgical coal, in our baseline and Panama Canal chokepoint scenario.}
    \label{tab:suptradepanamamet}
    \begin{threeparttable}
    \centering
    \scriptsize
    \begin{tabular}{|l|l|l|l|l|l|l|l|l|l|l|}
    \hline
        Supplier & Coal type & \multicolumn{3}{c}{Volume supplied (Mt)}  &  \multicolumn{3}{c}{Supply price (\$/t)} & \multicolumn{3}{c}{Revenue (M\$)}\\ \hline
        ~ & ~ & Base & Scen. & Diff. & Base & Scen. & Diff. & Base & Scen. & Diff. \\ \hline
        Australia & HCC & 148.3 & 148.3 & 0 & 155.78 & 155.78 & 0 & 23,102 & 23,102 & 0 \\ \hline
        Australia & SSCC & 19.75 & 19.39 & -0.36 & 71.73 & 71.66 & -0.07 & 1,417 & 1,389 & -28 \\ \hline
        Australia & PCI & 20.09 & 20.09 & 0 & 88.58 & 88.58 & 0 & 1,780 & 1,780 & 0 \\ \hline
        Australia & Total Met & 188.15 & 187.79 & -0.36 & 139.77 & 139.9 & 0.13 & 26,299 & 26,271 & -28 \\ \hline
        N-America & HCC & 81.01 & 81.01 & 0 & 141.4 & 141.4 & 0 & 11,455 & 11,455 & 0 \\ \hline
        N-America & SSCC & 0 & 0 & 0 & 0 & 0 & 0 & 0 & 0 & 0 \\ \hline
        N-America & PCI & 3.96 & 3.96 & 0 & 84.65 & 84.65 & 0 & 335 & 335 & 0 \\ \hline
        N-America & Total Met & 84.96 & 84.96 & 0 & 138.76 & 138.76 & 0 & 11,790 & 11,790 & 0 \\ \hline
        Mongolia & HCC & 43.9 & 43.9 & 0 & 136.74 & 136.74 & 0 & 6,003 & 6,003 & 0 \\ \hline
        Mongolia & SSCC & 12.24 & 12.24 & 0 & 25.92 & 25.92 & 0 & 317 & 317 & 0 \\ \hline
        Mongolia & PCI & 0 & 0 & 0 & 0 & 0 & 0 & 0 & 0 & 0 \\ \hline
        Mongolia & Total Met & 56.14 & 56.14 & 0 & 112.58 & 112.58 & 0 & 6,320 & 6,320 & 0 \\ \hline
        Russia & HCC & 18.48 & 18.48 & 0 & 93.11 & 93.11 & 0 & 1,720 & 1,720 & 0 \\ \hline
        Russia & SSCC & 6.83 & 7.19 & 0.37 & 34.06 & 34.06 & 0 & 232 & 245 & 12 \\ \hline
        Russia & PCI & 25.8 & 25.8 & 0 & 38.46 & 38.46 & 0 & 992 & 992 & 0 \\ \hline
        Russia  & Total Met & 51.1 & 51.47 & 0.37 & 57.63 & 57.46 & -0.17 & 2,945 & 2,957 & 12 \\ \hline
    \end{tabular}
        \begin{tablenotes}
      \scriptsize
      \item Note: Columns 'Base' report results for the baseline projection. Columns labelled `Scen.' report results for a closure of the Panama Canal. Columns labelled `Diff.' report the differnce between the two. Coal types separated into Hard Coking Coal (HCC), Semi-soft Coking Coal (SSCC), Pulverised Coal for Injection (PCI) and total metallurgical coal (sum of HCC, SSCC, PCI).
    \end{tablenotes}
\end{threeparttable}
\end{table}

\begin{table}[!ht]
    \caption{Trade volumes, costs to consumers, and revenue, for thermal and  metallurgical coal, in our baseline and Panama Canal chokepoint scenario.}
    \label{tab:suptradepanama}
    \begin{threeparttable}
    \centering
    \scriptsize
    \begin{tabular}{|l|l|l|l|l|l|l|l|l|l|l|}
    \hline
        Supplier & Coal type & \multicolumn{3}{c}{Volume supplied (Mt)}  &  \multicolumn{3}{c}{Supply price (\$/t)} & \multicolumn{3}{c}{Revenue (M\$)}\\ \hline
         ~ & ~ & Base & Scen. & Diff. & Base & Scen. & Diff. & Base & Scen. & Diff. \\ \hline
        Australia & Metallurgical & 188.152 & 187.601 & -0.551 & 139.77 & 139.7 & -0.07 & 26,299 & 26,208 & -91 \\ \hline
        Australia & Thermal & 220.627 & 220.627 & 0 & 68.23 & 68.23 & 0 & 15,052 & 15,052 & 0 \\ \hline
        N-America & Metallurgical & 84.246 & 84.246 & 0 & 137.7 & 137.7 & 0 & 11,601 & 11,601 & 0 \\ \hline
        N-America & Thermal & 42.632 & 42.632 & 0 & 60.18 & 60.18 & 0 & 2,566 & 2,566 & 0 \\ \hline
        Colombia & Metallurgical & 0 & 0 & 0 & 0 & 0 & 0 & 0 & 0 & 0 \\ \hline
        Colombia & Thermal & 80.72 & 80.72 & 0 & 57.82 & 57.82 & 0 & 4,668 & 4,668 & 0 \\ \hline
        Indonesia & Metallurgical & 2.612 & 2.612 & 0 & 79.6 & 79.6 & 0 & 208 & 208 & 0 \\ \hline
        Indonesia & Thermal & 480.912 & 482.988 & 2.077 & 60.6 & 60.59 & -0.01 & 29,144 & 29,264 & 120 \\ \hline
        Mongolia & Metallurgical & 56.137 & 56.137 & 0 & 112.58 & 112.58 & 0 & 6,320 & 6,320 & 0 \\ \hline
        Mongolia & Thermal & 3.99 & 3.99 & 0 & 22.64 & 22.64 & 0 & 90 & 90 & 0 \\ \hline
        RoW & Metallurgical & 5.853 & 5.853 & 0 & 113.79 & 113.79 & 0 & 666 & 666 & 0 \\ \hline
        RoW & Thermal & 86.017 & 84.585 & -1.431 & 56.49 & 55.79 & -0.7 & 4,859 & 4,719 & -140 \\ \hline
        Russia & Metallurgical & 51.1 & 51.651 & 0.551 & 57.63 & 57.39 & -0.24 & 2,945 & 2,964 & 19 \\ \hline
        Russia & Thermal & 139.35 & 138.299 & -1.051 & 38.24 & 38.25 & 0.01 & 5,328 & 5,290 & -39 \\ \hline
        China  & Metallurgical & 648.116 & 648.116 & 0 & 106.41 & 106.41 & 0 & 68,967 & 68,967 & 0 \\ \hline
        China  & Thermal & 3207.279 & 3207.718 & 0.438 & 52.13 & 52.13 & 0 & 167,190 & 167,221 & 31 \\ \hline
    \end{tabular}
        \begin{tablenotes}
      \scriptsize
      \item Note: Columns 'Base' report results for the baseline projection. Columns labelled `Scen.' report results for a closure of the Panama Canal. Columns labelled `Diff.' report the differnce between the two.
    \end{tablenotes}
\end{threeparttable}
\end{table}

\begin{figure}[H] % hbtp for here, top, bottom, next available page. H for really just here
    \centering
    \includegraphics[width=1.0\textwidth]{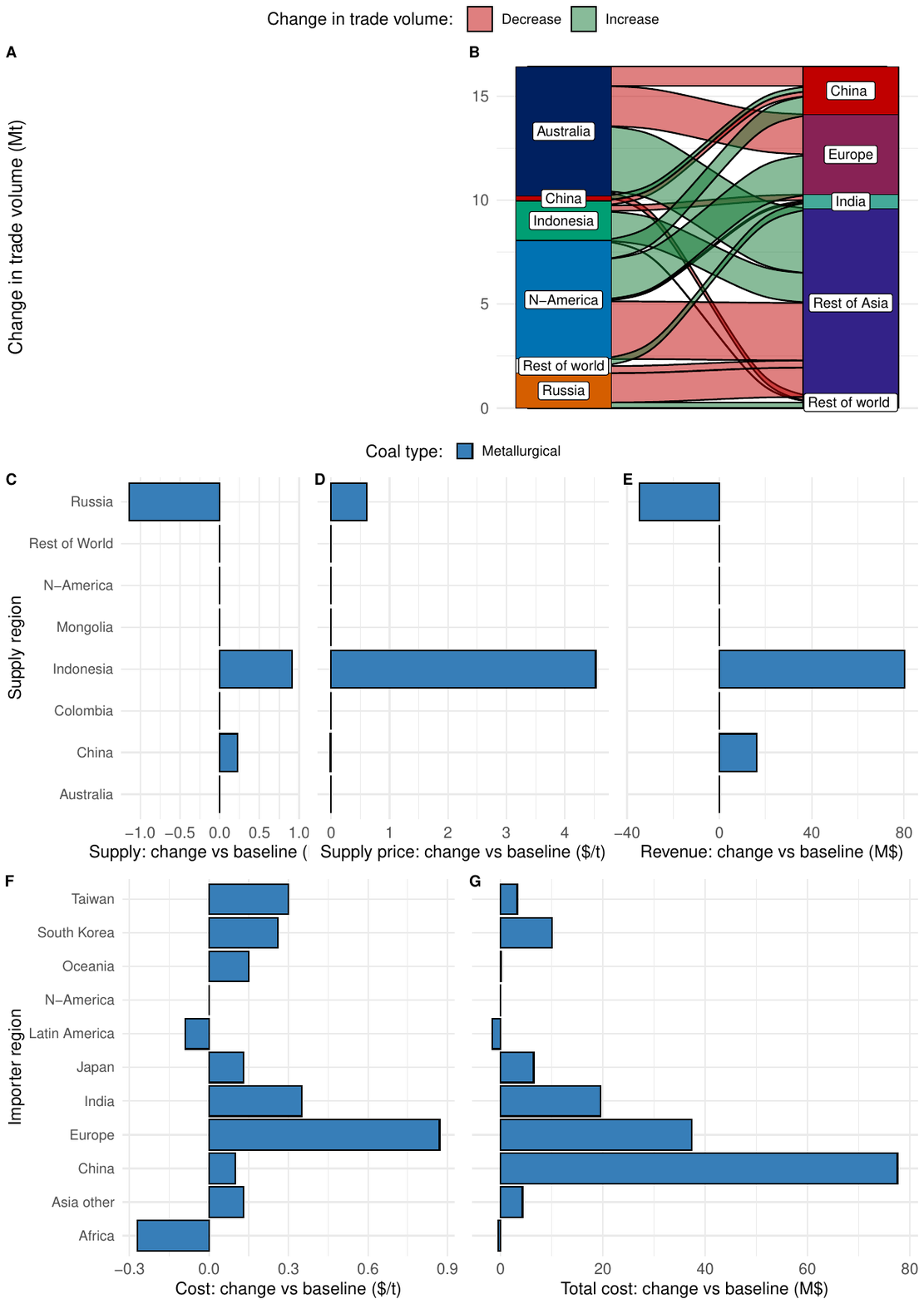}
        \caption{Changes to global coal trade in case of a restriction to maritime traffic through the Suez Canal, in the year 2020. 
        \scriptsize
        \textmd{Changes in volumes of for thermal coal (A) and metallurgical coal (B) traded between key exporter and importer country (groups), supply volume (C), price (D) and revenue (E) for exporters, and cost per ton (F) and total costs (G) for importers. Note: trade flow changes $<$1 Mt in panel A censored for legibility. We select to present results for the year 2020 here as trade flows of metallurgical coal in our baseline scenario for this year as the closest to the observed trade flows reported by Kpler 1, of as much as 49 Mt of metallurgical coal transported through this canal prior to the Red Sea crisis, “including from Australia to Europe (17 Mt), US East Coast to Asia (15 Mt, to India/China/Japan), and Western Russia to Asia (13 Mt, mostly to India)” (Supplementary Table \ref{tab:suptradepanamamet}). Note: there are no changes to thermal coal trade reported as our baseline scenario for 2020 projected no thermal coal to be transported via the Suez Canal in the year 2020 (Supplementary Table \ref{tab:suezmettrade}). Similarly to our original results (Fig. \ref{fig:suez}F in the main text), our model suggest cost increases of such a closure would be limited to below \$1/t for the most affected market}
        }
    \label{fig:suez2020}
\end{figure}

\begin{table}[!ht]
    \caption{Trade flows of coal through the Suez Canal in our baseline scenario, with metallurgical trade flows split by origin-destination pairs.}
    \label{tab:suezmettrade}
    \centering
    \scriptsize
    \begin{tabular}{|l|l|l|l|l|l|l|l|l|l|l|l|l|}
    \hline
         & 2015 & 2016 & 2017 & 2018 & 2019 & 2020 & 2021 & 2022 & 2023 & 2024 & 2025 & 2026 \\ \hline
        Thermal & 11.58 & 11.09 & 0 & 13.36 & 8.52 & 0 & 4.19 & 0 & 0 & 3.74 & 5.36 & 5.44 \\ \hline
        Metallurgical, of which: & 47.53 & 44.05 & 27.05 & 23.52 & 20.85 & 45.43 & 12.99 & 15.96 & 19.03 & 18.77 & 18.33 & 15 \\ \hline
        Australia to Europe & 30.41 & 32.38 & 21.29 & 9.31 & 8.46 & 25.12 & 0.51 & 0 & 0.01 & 0.01 & 0.01 & 0.01 \\ \hline
        N-America to Asia & 11.75 & 4.2 & 0.03 & 0 & 0 & 10.83 & 0 & 0 & 0 & 0 & 0 & 0 \\ \hline
        Russia West to Asia & 5.37 & 7.47 & 5.73 & 14.21 & 12.38 & 9.14 & 12.18 & 13.81 & 17.64 & 17.38 & 16.93 & 14.99 \\ \hline
        Other & 0 & 0 & 0 & 0 & 0 & 0.34 & 0.29 & 2.15 & 1.37 & 1.38 & 1.38 & 0 \\ \hline
    \end{tabular}
\end{table}

\begin{figure}[H] % hbtp for here, top, bottom, next available page. H for really just here
    \centering
    \includegraphics[width=1.0\textwidth]{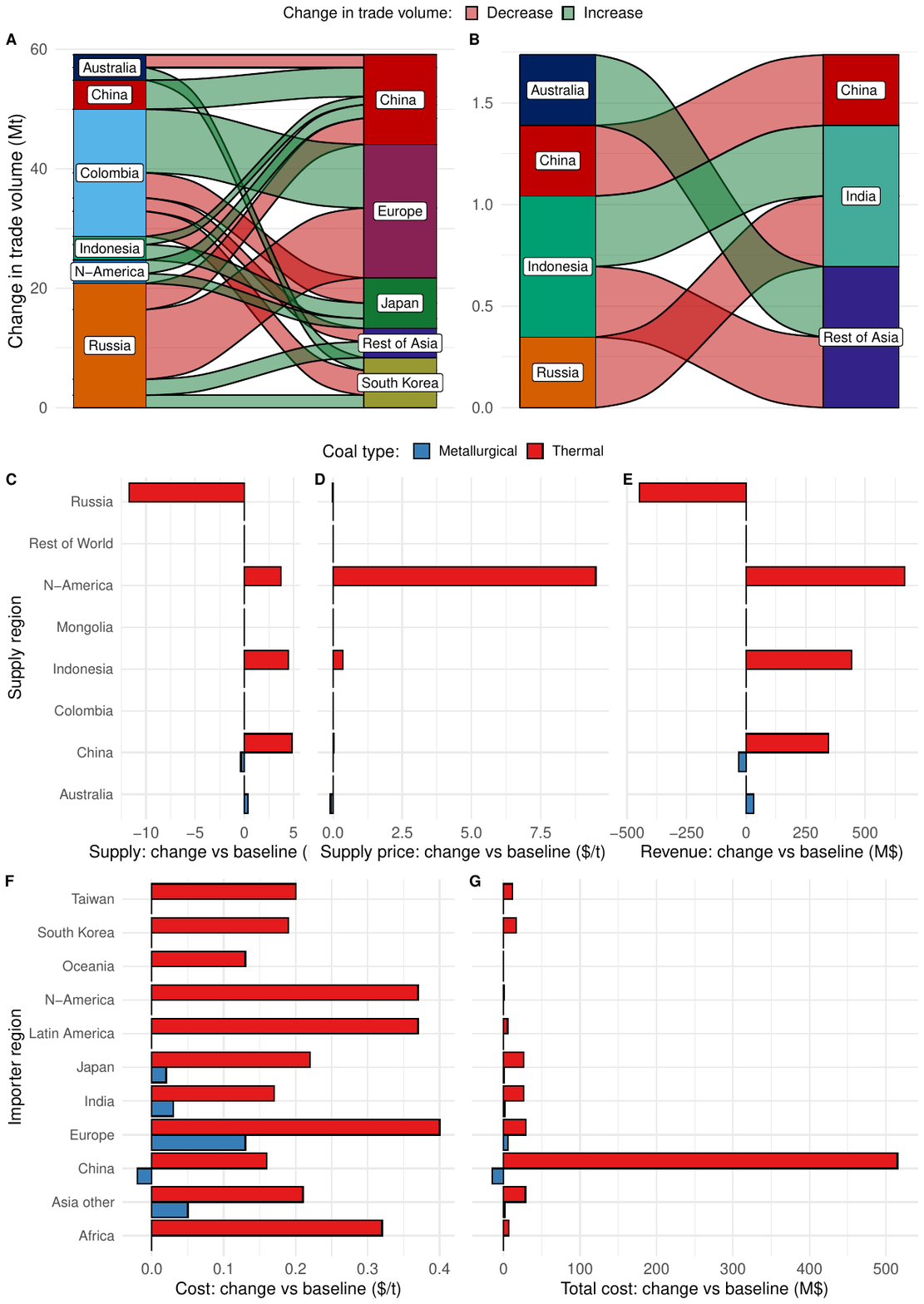}
        \caption{Changes to global coal trade in case of a restriction to maritime traffic through the Danish Strait. 
        \textmd{Changes in volumes of for thermal coal (A) and metallurgical coal (B) traded between key exporter and importer country (groups), supply volume (C), price (D) and revenue (E) for exporters, and cost per ton (F) and total costs (G) for importers. Note: trade flow changes $<$1 Mt in panel A censored for legibility}
        }
    \label{fig:danish}
\end{figure}

\begin{figure}[H] % hbtp for here, top, bottom, next available page. H for really just here
    \centering
    \includegraphics[width=1.0\textwidth]{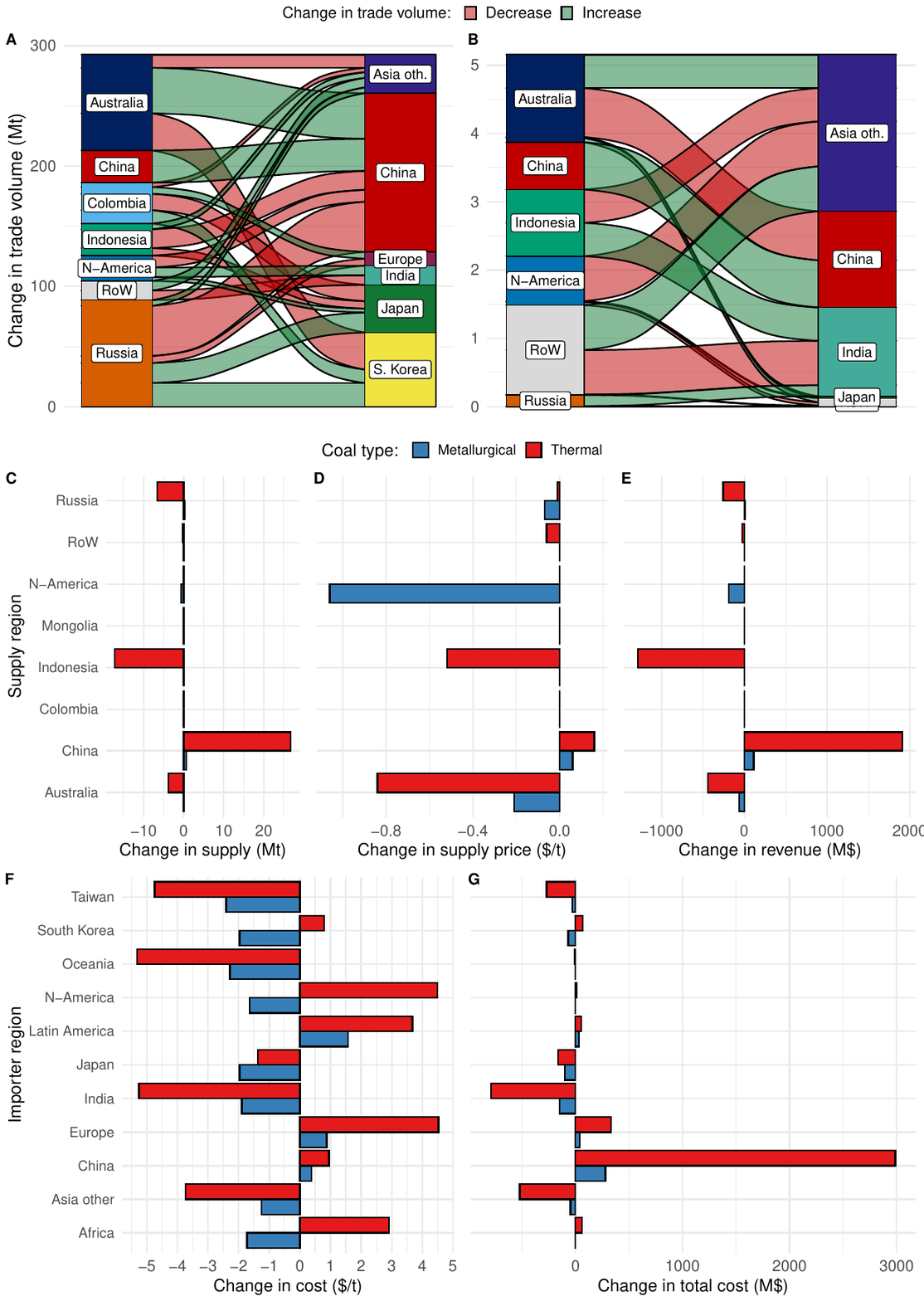}
        \caption{Changes to global coal trade in case of a restriction to maritime traffic in the South and East China Sea, but with ports in China’s northern provinces still capable of receiving seaborne imports. 
        \scriptsize
        \textmd{Changes in volumes of for thermal coal (A) and metallurgical coal (B) traded between key exporter and importer country (groups), supply volume (C), price (D) and revenue (E) for exporters, and cost per ton (F) and total costs (G) for importers. Note: trade flow changes $<$1 Mt in panel A censored for legibility. Note: trade flow changes $<$2 Mt censored in the plot for thermal coal for legibility. Note also that costs in North and Latin American, European, and African markets rise, as more of Russia’s coal is supplied into the Asian market, drawing away supplies into the Atlantic market, raising prices there}
        }
    \label{fig:SECSN}
\end{figure}

\begin{figure}[H] % hbtp for here, top, bottom, next available page. H for really just here
    \centering
    \includegraphics[width=1.0\textwidth]{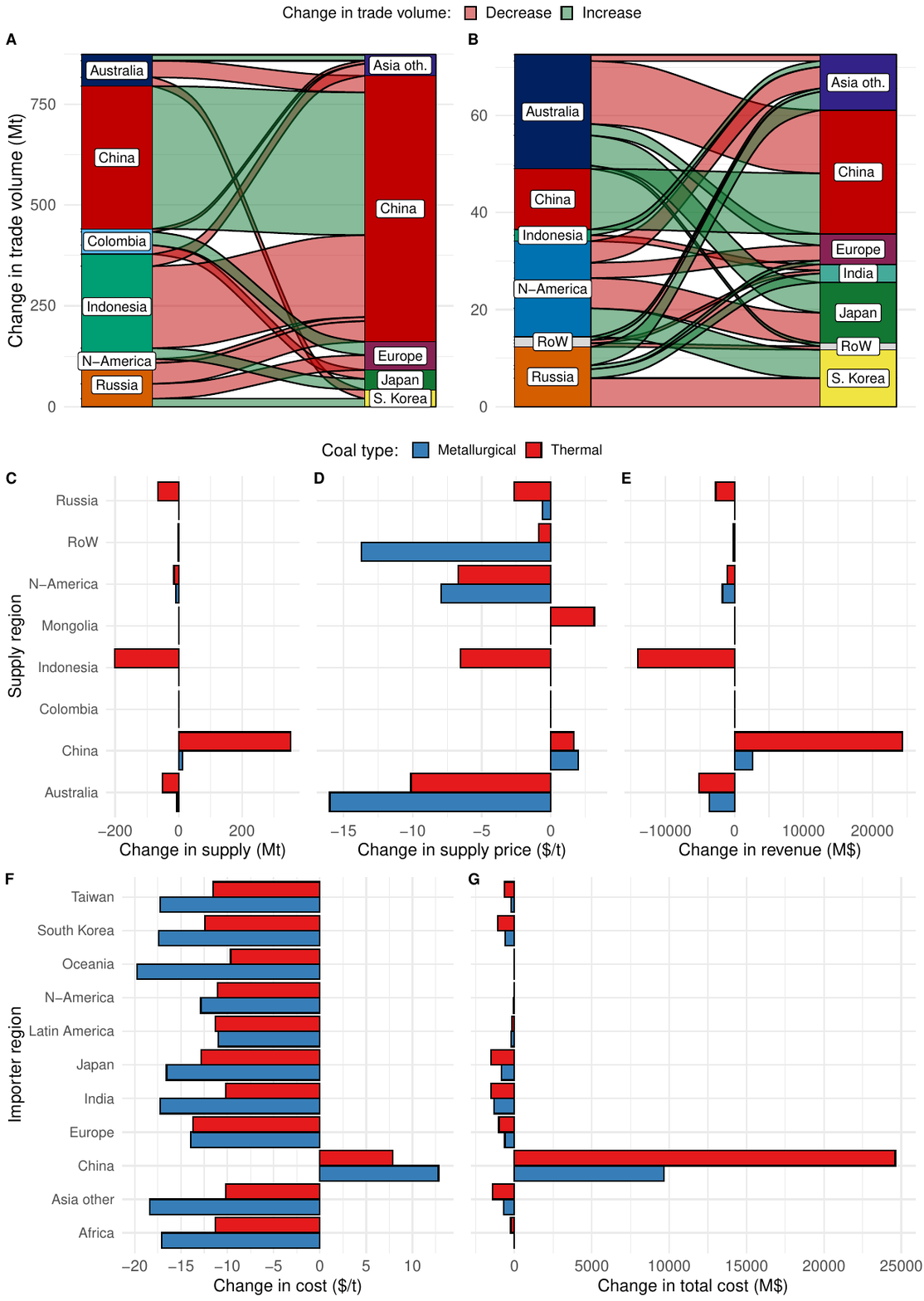}
        \caption{Changes to global coal trade in case of a restriction to maritime traffic through the South and East China Sea. 
        \scriptsize
        \textmd{Scenario for a restriction of Chinese seaborne imports to 50 Mt (approximately 12.5\% of baseline imports). Changes in volumes of for thermal coal (A) and metallurgical coal (B) traded between key exporter and importer country (groups), supply volume (C), price (D) and revenue (E) for exporters, and cost per ton (F) and total costs (G) for importers. Note: trade flow changes $<$6 Mt in panel A and changes $<$0.5 Mt in panel B censored for legibility}
        }
    \label{fig:SECS50}
\end{figure}

\begin{table}
\caption{Demand nodes, countries included, and seaborne import demand}
\label{tab:demandbynode}
\scriptsize
\centering
\begin{threeparttable}
\begin{tblr}{
  width = \linewidth,
  colspec = {Q[40]Q[70]Q[35]Q[35]Q[35]Q[35]},
  }
    \hline
        region & Countries included (ISO3 codes) & Imports thermal coal (Mt) & Imports metallurgical coal (Mt) & Imports thermal coal (\% of global) & Imports metallurgical coal (\% of global) \\ \hline
        Global & Global & 1225.5 & 391.1 & 100.00\% & 100.00\% \\ \hline
        China\textsuperscript{1,2} & CHN & 490.5 & 118.6 & 40.00\% & 30.30\% \\ \hline
        India & IND & 161.3 & 67.9 & 13.20\% & 17.40\% \\ \hline
        Japan & JPN & 134.9 & 53.7 & 11.00\% & 13.70\% \\ \hline
        South Korea & KOR & 94 & 33.6 & 7.70\% & 8.60\% \\ \hline
        EU\textsuperscript{1,3} & ALB, BEL, BGR, CYP, DEU, DNK, ESP, EST, FRA, GRC, HRV, ITA, LTU, LVA, NLD, POL, PRT, ROM, SVN & 91.5 & 38.8 & 7.50\% & 9.90\% \\ \hline
        South East Asia & BGD, BRN, KHM, LKA, MMR, MYS, SGP, THA, VNM & 85.8 & 16.1 & 7.00\% & 4.10\% \\ \hline
        Taiwan & TWN & 62.5 & 11 & 5.10\% & 2.80\% \\ \hline
        Turkey & TUR & 27.5 & 12.6 & 2.20\% & 3.20\% \\ \hline
        Northern Africa & DZA, EGY, LBY, MAR, SDN, TUN & 14.9 & 0.4 & 1.20\% & 0.10\% \\ \hline
        Southern Asia & IRN, MDV, PAK & 10.6 & 0.1 & 0.90\% & 0.00\% \\ \hline
        Indonesia & IDN & 9.7 & 6.3 & 0.80\% & 1.60\% \\ \hline
        Caribbean & ABW, BMU, BRB, CRI, CUB, CUW, DOM, GTM, HND, HTI, JAM, MEX, PAN, PRI, TTO & 7.9 & 1.2 & 0.60\% & 0.30\% \\ \hline
        Israel & ISR & 6.4 & 0 & 0.50\% & 0.00\% \\ \hline
        Brazil & BRA & 4 & 16.4 & 0.30\% & 4.20\% \\ \hline
        UK Ireland Iceland & GBR, IRL, ISL & 3.6 & 3.7 & 0.30\% & 0.90\% \\ \hline
        Eastern Africa & DJI, ERI, KEN, MDG, MOZ, MUS, SOM, TZA & 3.3 & 0.2 & 0.30\% & 0.10\% \\ \hline
        Western Asia & ARE, AZE, BHR, GEO, IRQ, JOR, KWT, LBN, OMN, QAT, SAU, SYR, YEM & 3 & 0.6 & 0.20\% & 0.20\% \\ \hline
        Scandinavia & FIN, NOR, SWE & 2.7 & 3.1 & 0.20\% & 0.80\% \\ \hline
        Canada East & CAN & 2.6 & 3 & 0.20\% & 0.80\% \\ \hline
        Western Africa & BEN, CIV, CPV, GHA, GIN, LBR, MRT, NGA, SEN, SLE, TGO & 1.9 & 0 & 0.20\% & 0.00\% \\ \hline
        Central America West & CRI, GTM, HND, MEX, NIC, PAN, SLV & 1.3 & 0.1 & 0.10\% & 0.00\% \\ \hline
        Melanesia & FJI, NCL, PNG, PYF, SLB & 1.1 & 0 & 0.10\% & 0.00\% \\ \hline
        New Zealand & NZL & 0.9 & 0 & 0.10\% & 0.00\% \\ \hline
        Ukraine & UKR & 0.8 & 1 & 0.10\% & 0.30\% \\ \hline
        Argentina Uruguay & ARG, URY & 0.7 & 1.1 & 0.10\% & 0.30\% \\ \hline
        South America West & COL, ECU, PER & 0.7 & 0.1 & 0.10\% & 0.00\% \\ \hline
        Middle Africa & AGO, CMR, COG, GAB, ZAR & 0.4 & 0.1 & 0.00\% & 0.00\% \\ \hline
        Australia & AUS & 0.4 & 0.4 & 0.00\% & 0.10\% \\ \hline
        Southern Africa & NAM, ZAF & 0.3 & 1 & 0.00\% & 0.30\% \\ \hline
        South America North & COL, GUY, SUR, VEN & 0.2 & 0 & 0.00\% & 0.00\% \\ \hline
        Russia East & RUS & 0.1 & 0 & 0.00\% & 0.00\% \\ \hline
        Canada West & CAN & 0 & 0 & 0.00\% & 0.00\% \\ \hline
        Chile & CHI & 0 & 0 & 0.00\% & 0.00\% \\ \hline
        Philippines & PHL & 0 & 0 & 0.00\% & 0.00\% \\ \hline
        US East & USA & 0 & 0 & 0.00\% & 0.00\% \\ \hline
        US West & USA & 0 & 0 & 0.00\% & 0.00\% \\ \hline
\end{tblr}
        \begin{tablenotes}
      \item Notes: demand is forecast demand for 2026; data for historical volumes of seaborne imports from Kpler\cite{Kpler2024Dry}, see methods section in the main manuscript for method of extrapolation to 2026; 1) includes landborne imports; 2) data source is SX Coal\cite{SX2026Coal}; 3) data source is IEA coal information\cite{IEA2026CoalInfo}.
    \end{tablenotes}
\end{threeparttable}
\end{table}

\begin{figure}[H] % hbtp for here, top, bottom, next available page. H for really just here
    \centering
    \includegraphics[width=1.0\textwidth]{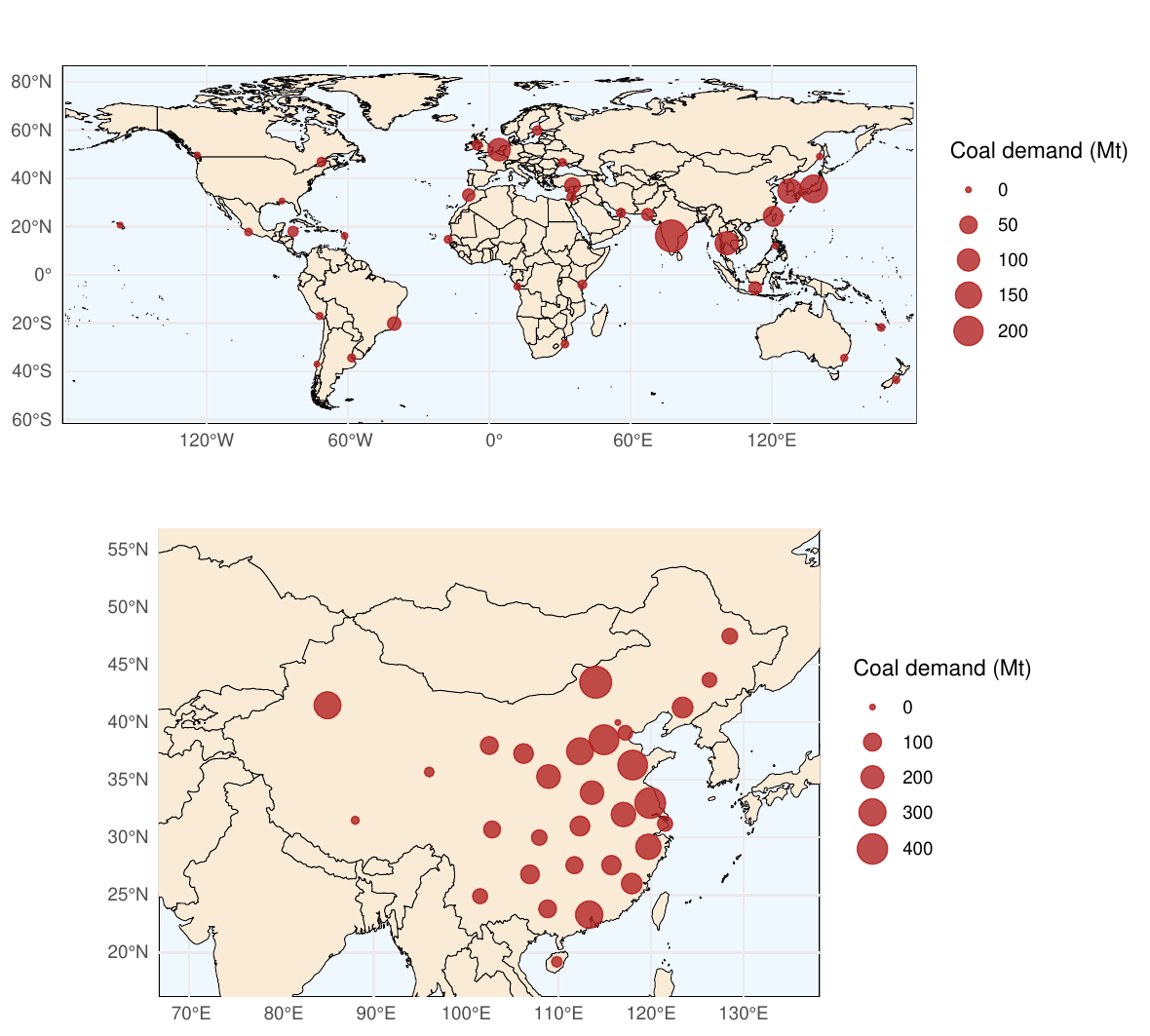}
        \caption{Location of demand nodes in our global network. 
        \textmd{For international nodes (top) and Chinese nodes (bottom). Note the size of the circle represents total volume of imports, i.e., thermal plus metallurgical coal. For Chinese nodes they represent total demand. Demand nodes are placed in the largest coal port in each country or country group (see Supplementary Table \ref{tab:demandbynode})}
        }
    \label{fig:demand}
\end{figure}

\newpage
\begin{figure}[H] % hbtp for here, top, bottom, next available page. H for really just here
    \centering
    \includegraphics[width=1.0\textwidth]{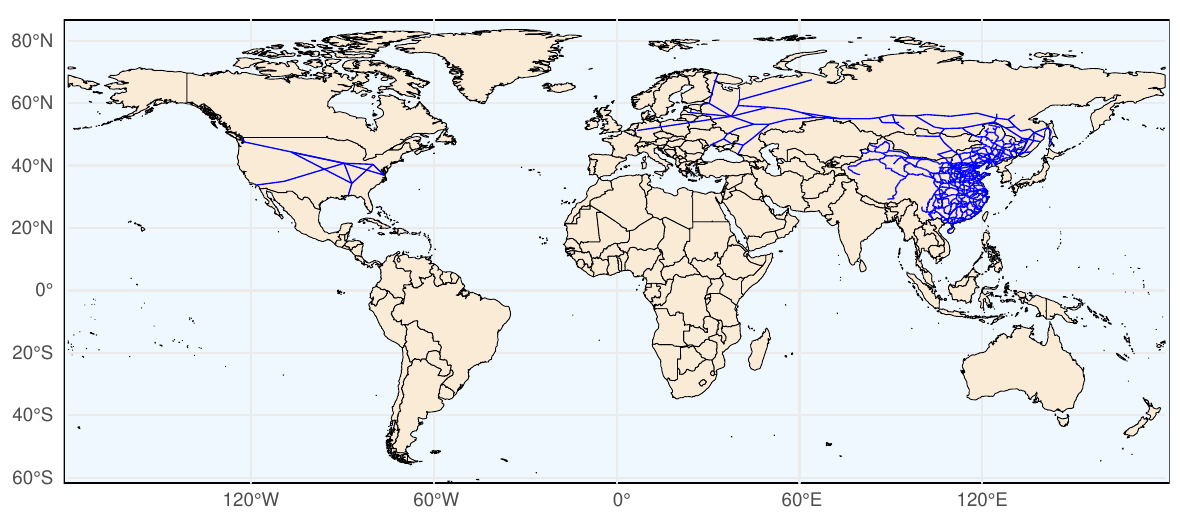}
        \caption{Rail networks in our global network. 
        \textmd{Representing networks in China, Mongolia, Russia, and the US.}
        }
    \label{fig:railnetwork}
\end{figure}

\newpage
\begin{figure}[H] % hbtp for here, top, bottom, next available page. H for really just here
    \centering
    \includegraphics[width=1.0\textwidth]{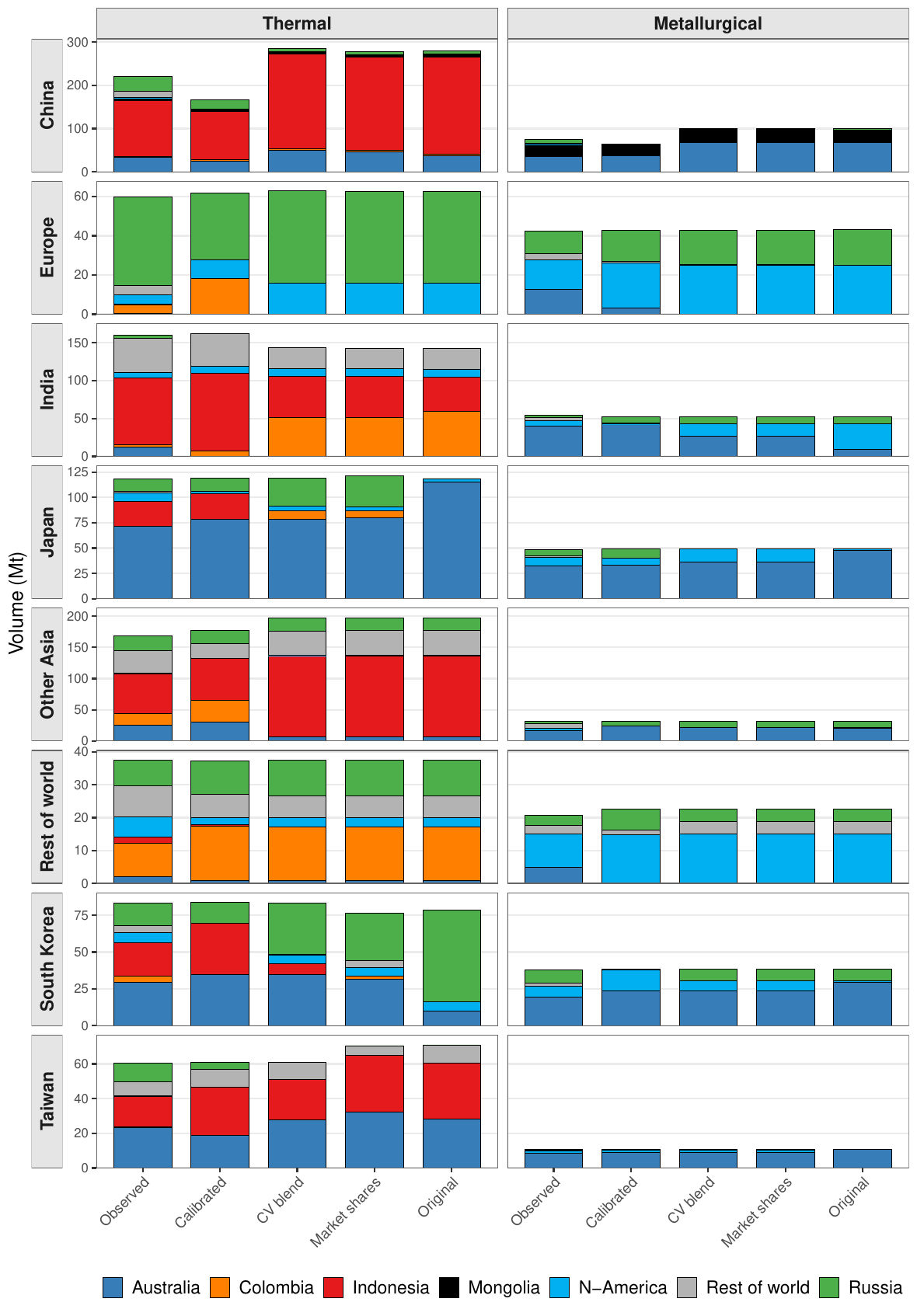}
        \caption{Observed and modelled origin of imports for country (group) pairs, at different calibration steps for the year 2019. 
        \textmd{‘Original’ = before any calibration; ‘Market shares’ = maximum exporter market shares to force import diversification; ‘CV blend’ = using a range of calorific value for the blend of thermal coal imported; ‘Calibrated’ = after applying preference parameters; ‘Observed’ = real world data.}
        }
    \label{fig:calibration}
\end{figure}

\newpage
\begin{figure}[H] % hbtp for here, top, bottom, next available page. H for really just here
    \centering
    \includegraphics[width=1.0\textwidth]{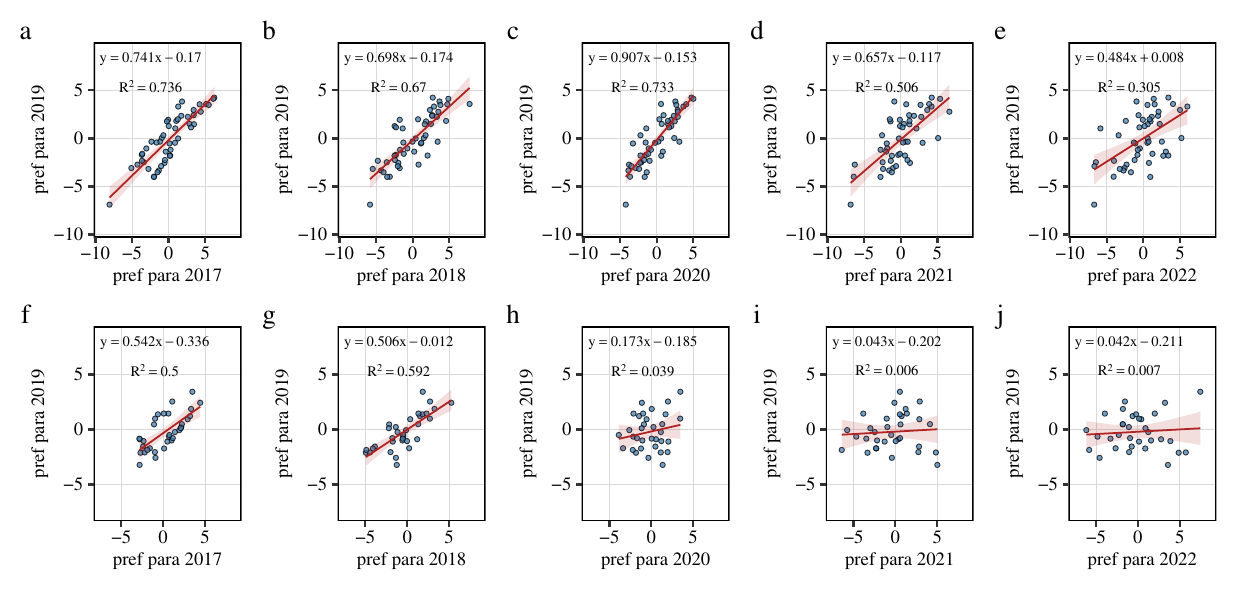}
        \caption{Correlation between preference parameters derived from different years of data. 
        \textmd{Each plotted against the preference parameters derived for 2019 (which are the ones used in our model), for thermal coal (panels a-e) and metallurgical coal (panels f-j). Each dot represents the preference parameter for a exporter-importer country (group) pair. The drop off in correlation with years 2020, 2021, and 2022 may indicate the effect of political embargoes and other issues disrupting cost-optimal logic driving global trade in coal.}
        }
    \label{fig:prefparacorrelation}
\end{figure}

\newpage
\begin{table}[!ht]
    \caption{Global imports of thermal coal split by countries that do or do not also have domestic coal production}
    \label{tab:importsthermal}
    \centering
    \scriptsize
    \begin{threeparttable}
    \begin{tabular}{|l|l|l|}
    \hline
        \multicolumn{3}{c}{Imports of thermal coal by countries that also have domestic production} \\ \hline
        Country (group) & Domestic production (Mt) & Imports (Mt) \\ \hline
        China & 3,847.10 & 371.9 \\ \hline
        India & 974.7 & 180 \\ \hline
        Vietnam & 51.9 & 55.3 \\ \hline
        Türkiye & 2.6 & 39.8 \\ \hline
        EU27 & 31 & 39.5 \\ \hline
        South, East and South-Eastern Asia & 855.7 & 100.8 \\ \hline
        Central Asia and the middle east & 108.7 & 2.2 \\ \hline
        Europe & 260.4 & 30.3 \\ \hline
        Africa & 253.1 & 2 \\ \hline
        Latin America and the Caribbean & 51.3 & 12.8 \\ \hline
        Northern America & 392.1 & 5.4 \\ \hline
        Oceania & 264.8 & 1.4 \\ \hline
        Total & 7,093.60 & 841.5 \\ \hline
        Total ex China & 3,246.50 & 469.6 \\ \hline
        China as \%age of global imports & ~ & 31.80\% \\ \hline
        Total excl. China as \%age of global imports & ~ & 40.20\% \\ \hline
        \multicolumn{3}{c}{} \\ \hline
        \multicolumn{3}{c}{Imports of thermal coal by countries that do not have domestic production} \\ \hline
        Country (group) & Domestic production (Mt) & Imports (Mt) \\ \hline
        Japan & ~ & 124.7 \\ \hline
        Korea & ~ & 86.4 \\ \hline
        Taiwan & ~ & 46.1 \\ \hline
        South, East and South-Eastern Asia & ~ & 32.3 \\ \hline
        Central Asia and the middle east & ~ & 6.1 \\ \hline
        Europe (other) & ~ & 2.8 \\ \hline
        Africa & ~ & 18.8 \\ \hline
        Latin America and the Caribbean & ~ & 9.7 \\ \hline
        Northern America & ~ & 0 \\ \hline
        Oceania & ~ & 0 \\ \hline
        Total & ~ & 326.8 \\ \hline
        Total as \%age of global imports & ~ & 28.00\% \\ \hline
    \end{tabular}
        \begin{tablenotes}
      \item Data source: IEA Coal Information 2026\cite{IEA2026CoalInfo}, data for the year 2025. Note volumes may not correspond with those listed in Supplementary Table \ref{tab:demandbynode} as the IEA and Kpler report on total imports vs seaborne imports, and may use different definitions for thermal and metallurgical coal.
    \end{tablenotes}
\end{threeparttable}
\end{table}

\begin{table}[!ht]
    \caption{Global imports of metallurgical coal split by countries that do or do not also have domestic coal production}
    \label{tab:importsmetcoal}
    \centering
    \scriptsize
    \begin{threeparttable}
    \begin{tabular}{|l|l|l|}
    \hline
        \multicolumn{3}{c}{Imports of thermal coal by countries that also have domestic production} \\ \hline
        Country (group) & Domestic production (Mt) & Imports (Mt) \\ \hline
        China & 593.2 & 118.6 \\ \hline
        India & 60.9 & 66.3 \\ \hline
        Indonesia & 7 & 26.3 \\ \hline
        Türkiye & 0.7 & 5.4 \\ \hline
        EU27 & 13 & 27.7 \\ \hline
        South, East and South-Eastern Asia & 80 & 0 \\ \hline
        Central Asia and the middle east & 5.1 & 6.1 \\ \hline
        Europe & 102.8 & 2.3 \\ \hline
        Africa & 12.2 & 0.5 \\ \hline
        Latin America and the Caribbean & 7.6 & 0.6 \\ \hline
        Northern America & 88.1 & 2.5 \\ \hline
        Oceania & 157 & 0.1 \\ \hline
        Total & 1,127.60 & 256.6 \\ \hline
        Total ex China & 534.4 & 138 \\ \hline
        China as \%age of global imports & ~ & 35.60\% \\ \hline
        Total ex China as \%age of global imports & ~ & 41.40\% \\ \hline
        \multicolumn{3}{c}{} \\ \hline
        \multicolumn{3}{c}{Imports of thermal coal by countries that do not have domestic production} \\ \hline
        Country (group) & Domestic production (Mt) & Imports (Mt) \\ \hline
        Japan & ~ & 35.9 \\ \hline
        Korea & ~ & 24.1 \\ \hline
        Taiwan & ~ & 5.4 \\ \hline
        Vietnam & ~ & 11 \\ \hline
        South, East and South-Eastern Asia & ~ & 0 \\ \hline
        Central Asia and the middle east & ~ & 0 \\ \hline
        Europe & ~ & 0 \\ \hline
        Africa & ~ & 0 \\ \hline
        Latin America and the Caribbean & ~ & 0 \\ \hline
        Northern America & ~ & 0 \\ \hline
        Oceania & ~ & 0 \\ \hline
        Total & ~ & 76.4 \\ \hline
        Total as \%age of global imports & ~ & 23.00\% \\ \hline
    \end{tabular}
        \begin{tablenotes}
      \item Data source: IEA Coal Information 2026\cite{IEA2026CoalInfo}, data for the year 2025. Note volumes may not correspond with those listed in Supplementary Table \ref{tab:demandbynode} as the IEA and Kpler report on total imports vs seaborne imports, and may use different definitions for thermal and metallurgical coal.
    \end{tablenotes}
\end{threeparttable}
\end{table}
\newpage

\end{document}